\documentclass{SciPost}

\usepackage[utf8]{inputenc}
\usepackage{booktabs}
\usepackage[multiple]{footmisc}
\usepackage{lipsum}
\usepackage{rotating}
\usepackage{perpage}
\usepackage{chronology}
\usepackage{amssymb}
\usepackage{amsbsy}
\usepackage{amsmath}
\usepackage{tikz}
\usepackage[T1]{fontenc}
\usepackage{etoolbox}
\usepackage{graphics}
\usepackage{abraces}
\usepackage{siunitx}
\usepackage{hyperref}
\usepackage{multirow}
\usepackage{mathtools}
\usepackage{bm}
\usepackage{url}
\usepackage{physics}
\usepackage{hyperref}
\usepackage[capitalise]{cleveref}
\usepackage{bbm}
\usepackage{color}
\usepackage{placeins}
\usepackage[normalem]{ulem}
\usepackage[compat=1.0.0]{tikz-feynman}
\usepackage{xparse}

\crefname{section}{Sec.}{subsections}
\Crefname{section}{Section}{Subsections}
\crefname{subsection}{Sec.}{subsections}
\Crefname{subsection}{Section}{Subsections}

\newcommand{\up}{\uparrow}
\newcommand{\down}{\downarrow}
\newcommand{\expec}[1]{\langle #1 \rangle}

\newcommand{\kf}{k_{\mathrm{F}}}
\newcommand{\Ef}{E_\mathrm{F}}

\newcommand{\dispersion}{\varepsilon}
\newcommand{\electronDispersion}{\dispersion_0}
\newcommand{\eFock}{\dispersion_\mathrm{Fock}}
\newcommand{\fockCorrection}{\dispersion_\mathrm{C}}

\newcommand{\debye}{\omega_\mathrm{D}}

\newcommand{\maxGap}{\Delta_\mathrm{max}}
\newcommand{\trueGap}{\Delta_\mathrm{true}}
\newcommand{\deltaC}{\Delta_\mathrm{C}}
\newcommand{\deltaPh}{\Delta_\mathrm{Ph}}

\newcommand{\volume}{\mathcal{V}}
\renewcommand{\vec}[1]{\bm{#1}}

\newcommand{\vk}{\vec{k}}
\newcommand{\vq}{\vec{q}}

\newcommand{\bonusModes}{secondary modes}
\newcommand{\BCSinteraction}{BCS interaction}
\newcommand{\CUTinteraction}{CUT interaction}

\NewDocumentCommand\ladder{O{}mm}{%
  c_{%
    \processmomenta{#2},%
    #3%
  }^{%
    \IfNoValueTF{{#1}}{{\phantom{\dagger}}}{#1}%
  }%
}

\ExplSyntaxOn
\newcommand{\processmomenta}[1]{%
  \tl_set:Nn \l_tmpa_tl {#1}%
  \tl_replace_all:Nnn \l_tmpa_tl {k}{\vk}%
  \tl_replace_all:Nnn \l_tmpa_tl {q}{\vq}%
  \tl_use:N \l_tmpa_tl%
}
\ExplSyntaxOff

\newcommand{\opSC}[1]{\ladder{-#1}{\down} \ladder{#1}{\up}}

\newcommand{\opf}[2][{\phantom{\dagger}}]{f_{#2}^{#1}}
\newcommand{\opn}[2]{n_{#1,#2}}

\newcommand{\greenSymbol}{\mathcal{G}}
\newcommand{\greens}[1]{\greenSymbol_\mathrm{#1} (\omega)}
\newcommand{\spectralSymbol}{\mathcal{A}}
\newcommand{\spectral}[1]{\spectralSymbol_\mathrm{#1}  (\omega)}
\newcommand{\im}{\mathrm{i}}

\newcommand{\bs}{\begin{subequations}}
\newcommand{\es}{\end{subequations}}
\newcommand{\be}{\begin{equation}}
\newcommand{\ee}{\end{equation}}

\newcommand{\edited}[1]{{#1}}

\def\hmath$#1${\texorpdfstring{{\rmfamily\mathrmit{#1}}}{#1}}

\def\plusheight{-\the\dimexpr\fontdimen22\textfont2\relax}

\hypersetup{
    colorlinks,
    linkcolor={red!50!black},
    citecolor={blue!50!black},
    urlcolor={blue!80!black}
}

\usepackage[bitstream-charter]{mathdesign}
\DeclareSymbolFont{usualmathcal}{OMS}{cmsy}{m}{n}
\DeclareSymbolFontAlphabet{\mathcal}{usualmathcal}

\fancypagestyle{SPstyle}{
\fancyhf{}
\lhead{\colorbox{scipostblue}{\bf \color{white} ~SciPost Physics }}
\rhead{{\bf \color{scipostdeepblue} ~Submission }}

\fancyfoot[C]{\textbf{\thepage}}
}

\begin{document}

\pagestyle{SPstyle}

\begin{center}{\Large \textbf{\color{scipostdeepblue}{
%%%%%%%%%% TODO: Write your article's title here
Collective modes in superconductors including Coulomb repulsion
%%%%%%%%%% END TODO: TITLE
}}}\end{center}

\begin{center}\textbf{
%%%%%%%%%% TODO: AUTHORS
% Write the author list here. 
% Use (full) first name (+ middle name initials) + surname format.
% Separate subsequent authors by a comma, omit comma and use "and" for the last author.
% Mark the corresponding author(s) with a superscript symbol in this order
% \star, \dagger, \ddagger, \circ, \S, \P, \parallel, ...
Joshua Alth\"user\textsuperscript{1$\star$} and
G\"otz S.~Uhrig\textsuperscript{1$\dagger$}
%%%%%%%%%% END TODO: AUTHORS
}\end{center}

\begin{center}
%%%%%%%%%% TODO: AFFILIATIONS
% Write all affiliations here.
% Format: institute, city, country
{\bf 1} Condensed Matter Theory, TU Dortmund University, Otto-Hahn Stra\ss{}e 4, 44227 Dortmund, Germany
%%%%%%%%%% END TODO: AFFILIATIONS
%%%%%%%%%% TODO: EMAIL
% Provide email address of corresponding author(s)
\\[\baselineskip]
$\star$ \href{mailto:joshua.althueser@tu-dortmund.de}{\small joshua.althueser@tu-dortmund.de}\,,\quad
$\dagger$ \href{mailto:goetz.uhrig@tu-dortmund.de}{\small goetz.uhrig@tu-dortmund.de}
%%%%%%%%%% END TODO: EMAIL
\end{center}

\section*{\color{scipostdeepblue}{Abstract}}
\textbf{\boldmath{%
  We numerically study the collective excitations present in isotropic superconductors including a screened Coulomb interaction.
  By varying the screening strength, we analyze its impact on the system.
  We use a formulation of the effective phonon-mediated interaction between electrons that depends on the energy transfer between particles,
  rather than being a constant in a small energy shell around the Fermi edge.
  This justifies considering also rather large attractive interactions.
  We compute the system's Green's functions using the iterated equations of motion (iEoM) approach,
  which ultimately enables a quantitative analysis of collective excitations.
  For weak couplings, we identify the well-known amplitude (Higgs) mode at the two-particle continuum's lower edge
  and the phase (Anderson-Bogoliubov) mode at $\omega = 0$ for a neutral system, which shifts to higher energies as the Coulomb interaction is switched on.
  As the phononic coupling is increased, the Higgs mode separates from the continuum, and additional phase and amplitude modes appear,
  persisting even in the presence Coulomb interactions.
}}

\vspace{\baselineskip}

%%%%%%%%%% BLOCK: Copyright information
% This block will be filled during the proof stage, and finilized just before publication.
% It exists here only as a placeholder, and should not be modified by authors.
\noindent\textcolor{white!90!black}{%
\fbox{\parbox{0.975\linewidth}{%
\textcolor{white!40!black}{\begin{tabular}{lr}%
  \begin{minipage}{0.6\textwidth}%
    {\small Copyright attribution to authors. \newline
    This work is a submission to SciPost Physics. \newline
    License information to appear upon publication. \newline
    Publication information to appear upon publication.}
  \end{minipage} & \begin{minipage}{0.4\textwidth}
    {\small Received Date \newline Accepted Date \newline Published Date}%
  \end{minipage}
\end{tabular}}
}}
}
%%%%%%%%%% BLOCK: Copyright information

%%%%%%%%%% TODO: LINENO
% For convenience during refereeing we turn on line numbers:
% \linenumbers
% You should run LaTeX twice in order for the line numbers to appear.
%%%%%%%%%% END TODO: LINENO

%%%%%%%%%% TODO: TOC 
% Guideline: if your paper is longer that 6 pages, include a TOC
% To remove the TOC, simply cut the following block
\vspace{10pt}
\noindent\rule{\textwidth}{1pt}
\tableofcontents
\noindent\rule{\textwidth}{1pt}
\vspace{10pt}
%%%%%%%%%% END TODO: TOC

%%%%%%%%% TODO: CONTENTS 
% Write your article contents here, starting from first \section.
% An example structure is given below.

%%%%%%%%%%%%%%%%%%%%%%%%%%%%%%%%%%%%%%%%%%%%%%%%%%%%%%%%%%%%%%%%%%%%%%%%%%%%%%%%%%%%%%%%%%%%%%%%%%%%%%%%%%%%%%%%%%%%%
%%%%%%%%%%%%%%%%%%%%%%%%%%%%%%%%%%%%%%%%%%%%%%%%%%%%%%%%%%%%%%%%%%%%%%%%%%%%%%%%%%%%%%%%%%%%%%%%%%%%%%%%%%%%%%%%%%%%%
%%%%%                                               Introduction                                                %%%%%
%%%%%%%%%%%%%%%%%%%%%%%%%%%%%%%%%%%%%%%%%%%%%%%%%%%%%%%%%%%%%%%%%%%%%%%%%%%%%%%%%%%%%%%%%%%%%%%%%%%%%%%%%%%%%%%%%%%%%
%%%%%%%%%%%%%%%%%%%%%%%%%%%%%%%%%%%%%%%%%%%%%%%%%%%%%%%%%%%%%%%%%%%%%%%%%%%%%%%%%%%%%%%%%%%%%%%%%%%%%%%%%%%%%%%%%%%%%

\section{Introduction}\label{sec:introduction}

Ever since its discovery over a century ago, superconductivity has captivated researchers due to its unique properties.
The experimental observations defy any classical expectations by demonstrating perfect conductivity and diamagnetism.
Especially the former effect offers a myriad of possible practical applications, spurring this ever-expanding field of research.

Our focus is on enhancing our understanding of collective excitations in superconducting systems. 
While the general effect of the Coulomb interaction is well understood, we seek to address the following key questions in this paper.
How do collective excitations respond to screening effects? 
What relevant effects exist within the BCS channel? 
How does employing a more elaborate attractive interaction than standard BCS theory alter outcomes?
How do stronger attractions affect collective excitations?

The first successful theoretical descriptions utilized an isotropic interaction that is attractive near the Fermi edge but vanishes elsewhere \cite{bardeen1957,bogoliubov1958}.
Within this framework, two types of collective excitations emerge.
One corresponds to amplitude fluctuations of the order parameter (Higgs mode) while the other to its phase fluctuations (Anderson-Bogoliubov mode).
The former is located at the lower edge of the two-particle continuum. 
It has been observed both experimentally and theoretically by a plethora of previous studies 
\cite{kulik1981,volkov1973,schmid1975,varma2002,yuzbashyan2006,cea2014,measson2014,tsuji2015,krull2016,reinhoffer2022,sulaiman2024,fischer2018,schwarz2020,dzero2024,althuser2024}. 
Its excitation energy is widely interpreted as the minimum energy necessary to break up a Cooper pair.

In the absence of long-range Coulomb interactions, i.e., in a neutral superfluid, the phase mode becomes a gapless Goldstone mode.
The inclusion of the Coulomb interaction, however, couples the phase of the order parameter to the electromagnetic field, 
shifting it toward the plasma frequency according to the common lore \cite{anderson1958,bogoliubov1958,brieskorn1974,schmid1975,simanek1975,schon1976,kulik1981,maiti2015,fischer2018,sun2020,fan2022,althuser2024}.

In this article, we numerically investigate superconducting systems.
Omitting the Coulomb interaction and assuming a constant attractive interaction around the Fermi edge, 
we confirm the expected results exactly.
We extend this analysis by incorporating an energy-dependent interaction and subsequently including Coulomb effects.

The attraction between two electrons in a conventional superconductor is typically attributed to an electron-phonon interaction 
that can be reinterpreted as an effective electron-electron interaction.
Fr\"ohlich presented the first description of this mechanism. 
While his result explains attraction at small energy transfers, it becomes repulsive at larger ones and exhibits linear divergences \cite{frohlich1952}.
Subsequent \textsl{ans\"atze} provided alternative descriptions with more favorable characteristics, 
in that they are for no parameter regime repulsive and are confined to a small energy region \cite{lenz1996,mielke1997,mielke1997a,hubsch2003,kehrein2006,krull2012}.
While all these distinct formulations differ in their description of processes that have a finite energy transfer,
they are identical for all real, i.e., on-shell, processes.
Naturally, this has to be the case because these processes can be measured, at least in principle \cite{kehrein2006}.

We adopt a result from a flow-equation approach, namely, a continuous unitary transformation (CUT) \cite{mielke1997,mielke1997a,krull2012,schmiedinghoff2022,walther2023,hering2024}.
In the BCS channel, this formulation yields an attractive interaction proportional to a Heaviside function that depends on the magnitude of the energy transfer \cite{krull2012}.
Thus, the interaction is attractive for small energy-transfers and vanishes otherwise.
In this regard, it is better suited for the description of superconducting system's than Fr\"ohlich's result as it encompasses neither repulsive regimes nor divergences in the BCS channel.

To compute the momentum-dependent superconducting order parameter we use a mean-field approximation to decouple the interaction terms.
This yields gaps qualitatively and quantitatively similar to those predicted by standard BCS theory \cite{bardeen1957,bogoliubov1958},
yet lacks the sharp cutoff observed at a certain distance from the Fermi edge.
For small interaction strengths, computing the collective excitations using this interaction yields qualitatively the same results as a constant interaction.
However, stronger interaction causes unexpected modes beyond the standard amplitude and phase mode to emerge from the two-particle continuum.

The Eliashberg formalism represents an alternative approach to superconductivity.
Here, the retardation effects are included explicitly \cite{eliashberg1960,combescot1990,joas2002,chang2007}.
These are commonly used to argue why the phononic attraction dominates over the instantaneous Coulomb repulsion.
In that regard, the basis transformation which we rely on in the derivation of the attractive interaction provides another advantage over Fr\"ohlich's expression.
Namely, it vanishes if the energy transfer is larger than the phonon energy $\omega_\mathrm{D}$.
Consequently, only scattering processes on a timescale larger than $1 / \omega_\mathrm{D}$ contribute.
This is precisely what is meant by retardation.
Hence, the employed formalism captures the essential effects of retardation \cite{kehrein2006}.

Additionally, we incorporate the Coulomb interaction to study its impact on both the order parameter and the collective excitations.
At the mean-field level, the Coulomb interaction effectively renormalizes the attractive interaction by a pseudopotential $\mu^*$ \cite{bogoliubov1958,morel1962}
and causes the gap function to extend to arbitrary momenta.
The order parameter also switches its sign close to the Fermi edge to avoid the repulsive nature of the Coulomb interaction as long as possible \cite{morel1962,carbotte1990,sigrist2005}.
Investigating the collective excitations, we see the primary phase mode shifting towards high energies in accordance with expectations.
This behavior can be continuously tracked by varying the screening strength of the Coulomb interaction.
The aforementioned additional modes persist even after switching on the Coulomb repulsion.

Studying collective excitations lies beyond a simple mean-field approach.
Therefore, we turn to the iterated equations of motion (iEoM) approach.
Conceptually, this method begins with the Heisenberg equations of motion and selects a suitable operator basis,
which is extended to represent all terms arising from commutation with the Hamiltonian.
Naturally, this leads to an infinite hierarchy of equations that must be truncated for practical computations.

This procedure has been successfully applied in prior studies 
including comparisons with density matrix formalism for the quantum Rabi model \cite{kalthoff2017}
and applications to interaction quenches as well as collective excitations in Hubbard models \cite{althuser2024,uhrig2009,hamerla2013,hamerla2014}.
In this article, we will follow the procedure outlined in Ref. \cite{althuser2024} to compute the Green's functions of the system and its collective excitations.
Lifetime effects of the quasiparticles are neglected by this procedure.
Nevertheless, due to the energy gap in the superconducting phase, lifetime effects are suppressed at low energies.

The remainder of the article is organized as follows:
We briefly introduce the model under study in \cref{sec:model}.
In \cref{sec:mean_field}, we show our mean-field calculations and discuss the results.
The study of the collective excitations is presented in \cref{sec:mode_analysis}.
Finally, we summarize the results, conclude, and provide an outlook in \cref{sec:conclusion}.

%%%%%%%%%%%%%%%%%%%%%%%%%%%%%%%%%%%%%%%%%%%%%%%%%%%%%%%%%%%%%%%%%%%%%%%%%%%%%%%%%%%%%%%%%%%%%%%%%%%%%%%%%%%%%%%%%%%%%
%%%%%%%%%%%%%%%%%%%%%%%%%%%%%%%%%%%%%%%%%%%%%%%%%%%%%%%%%%%%%%%%%%%%%%%%%%%%%%%%%%%%%%%%%%%%%%%%%%%%%%%%%%%%%%%%%%%%%
%%%%%                                                   Model                                                   %%%%%
%%%%%%%%%%%%%%%%%%%%%%%%%%%%%%%%%%%%%%%%%%%%%%%%%%%%%%%%%%%%%%%%%%%%%%%%%%%%%%%%%%%%%%%%%%%%%%%%%%%%%%%%%%%%%%%%%%%%%
%%%%%%%%%%%%%%%%%%%%%%%%%%%%%%%%%%%%%%%%%%%%%%%%%%%%%%%%%%%%%%%%%%%%%%%%%%%%%%%%%%%%%%%%%%%%%%%%%%%%%%%%%%%%%%%%%%%%%

\section{Model}\label{sec:model}

\begin{figure}
  \centering
  \includegraphics{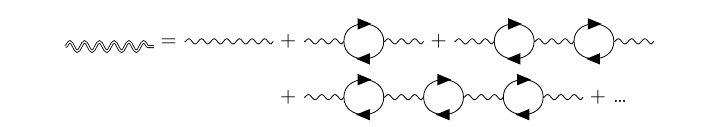}
  \caption{Diagramatic representation of the effective screened Coulomb interaction.}
  \label{fig:feynman}
\end{figure}

We focus on a rather general Hamiltonian at zero temperature given by
\begin{equation}
\label{eqn:base_hamiltonian}
  H = H_\mathrm{kin} + H_\mathrm{Ph} + H_\mathrm{C} + H_\mathrm{BG},
\end{equation}
where $H_\mathrm{kin}$ describes the kinetic part,
$H_\mathrm{Ph}$ describes an effective electron-electron interaction mediated by phonons, 
$H_\mathrm{C}$ describes the repulsive Coulomb interaction between electrons,
and $H_\mathrm{BG}$ describes the electronic interaction between the electrons and the atomic nuclei,
which we represent as a uniform positive background charge $\rho$.
This term exactly cancels with the divergent Hartree contribution in $H_\mathrm{C}$ \cite{rickayzen1980,czycholl2008}.
In the remainder of this article, we will refer to contributions due to $H_\mathrm{Ph}$ as \emph{phononic}.
The individual terms are given by
\bs
\begin{align}
  H_\mathrm{Kin} &= \sum_{\vk \sigma} \electronDispersion (\vk) \ladder[\dagger]{k}{\sigma} \ladder{k}{\sigma} \\
\label{eqn:phonon_hamiltonian}
  H_\mathrm{Ph} &= \frac{1}{\volume} \sum_{\vk \vk' \sigma} g(\vk, \vk') 
      \ladder[\dagger]{k}{\sigma} \ladder[\dagger]{-k}{-\sigma} \ladder{-k'}{-\sigma} \ladder{k'}{\sigma} \\
\label{eqn:coulomb_hamiltonian}
  H_\mathrm{C} &= \frac{1}{2\volume} \sum_{\substack{ \vk \vk' \vq \\ \sigma \sigma'}} V(|\vq|) 
      \ladder[\dagger]{k}{\sigma} \ladder[\dagger]{k'}{\sigma'} \ladder{k'-q}{\sigma'} \ladder{k+q}{\sigma} \\
\label{eqn:background_hamiltonian}
  H_\mathrm{BG} &= - \frac{1}{2} \sum_{ \vk \sigma } V(0) \rho \ladder[\dagger]{k}{\sigma} \ladder{k}{\sigma}.
\end{align}
\es
Here, $\volume$ is the system's volume, $\electronDispersion (k) = \hbar^2 k^2 / (2 m_e) - \Ef$ the free particle dispersion, 
$\Ef$ the Fermi energy, $\hbar$ the reduced Planck's constant, and $m_e$ the electron mass.
The operator $\ladder[(\dagger)]{k}{\sigma}$ annihilates (creates) an electron with momentum $\vk$ and spin $\sigma$.
The Coulomb potential is given by
\begin{equation}
  V(q) = \frac{e^2}{\epsilon_0} \frac{1}{q^2 + k_s^2},
\end{equation}
where $e$ is the elementary charge, $\epsilon_0$ the vacuum permittivity, and $k_s$ the Thomas-Fermi screening wavevector.
\edited{This is a standard procedure to capture the electronic screening effects in metals that result from diagrams as depicted in \cref{fig:feynman} \cite{rickayzen1980,simonato2023}.
Moreover, this choice of potential provides a mathematical regularization to avoid divisions by zero.
Often, the Coulomb interaction is considered to be reduced by a logarithm, cf. the theories presented by Tolmachev, and Morel and Anderson \cite{tolmachev1961,morel1962}.
As we present later, our method reproduces these results on the mean-field level for typical values of $k_s$.
The case of unscreened Coulomb interaction is retrieved in the limit $k_s \to 0$ wherever this limit can be taken.}

For comparison, we will also consider the Coulomb interaction only in the BCS channel
\begin{equation}
  \label{eqn:bcs_coulomb_hamiltonian}
  H_\mathrm{C}^{\text{(BCS)}} = \frac{1}{2\volume} \sum_{\vk \vk' \sigma} V(|\vk - \vk'|)
      \ladder[\dagger]{k}{\sigma} \ladder[\dagger]{-k}{-\sigma} \ladder{-k'}{-\sigma} \ladder{k'}{\sigma}.
\end{equation}
In this case, the background term is omitted as well because the Hartree contribution of the above term vanishes.

Obtaining an effective electron-electron interaction based on the physical electron-phonon interaction has been a major subject of past studies.
Many \textsl{ans\"atze} have been proposed that serve this purpose.
Moreover, the resulting effective interactions are not necessarily identical.
For motivation, let $S$ be the generator of the unitary transformation that decouples the phononic and electronic subsystems in $\mathcal{O}(M^2)$.
Then, one can add any interaction term in $\mathcal{O}(M^2)$ that conserves the number of phonons to $S$ while still achieving decoupling of the subsystems \cite{kehrein2006}.
One of the best-known formulations was presented by Fr\"ohlich \cite{frohlich1952}
\begin{equation}
  g^\text{Fr\"ohlich} (\vk,\vk') = \frac{|M_{\vk' - \vk}|^2 \omega_{\vk' - \vk}}{(\dispersion(\vk') - \dispersion(\vk))^2 - \omega_{\vk' - \vk}^2},
\end{equation}
where $M_{\vq}$ is the electron-phonon coupling strength,
$\omega_{\vq}$ is the phonon frequency, and $\dispersion(\vk)$ represents the single-particle energy.
This form has two significant disadvantages compared to other formulations, namely, that it is only attractive in a small energy interval and that it is divergent at $|\omega_{\vk' - \vk}| = |\dispersion(\vk' ) - \dispersion(\vk )|$.

Alternative \textsl{ans\"atze} resulted in expressions that remedied these issues by employing alternative basis transformations, see, e.g., Refs. \cite{kehrein2006,lenz1996,mielke1997,mielke1997a,hubsch2003,krull2012}.
The most important commonality among all these results is that they coincide for real, i.e., energy-conserving, processes.
This agreement is essential since these processes are the ones that are observable experimentally, at least in principle \cite{kehrein2006}.
In contrast, the virtual processes, i.e., the processes which do not conserve the total energy change depending on the specific basis transformation,
may be altered by different unitary transformations. 
This freedom of choice is successfully exploited in continuous unitary transformations which we employ here to derive the effective interaction in the BCS channel \cite{krull2012}
\begin{equation}
  g(\vk, \vk') = - \frac{|M_{\vk' - \vk}|^2}{\omega_{\vk' - \vk}} \Theta( \omega_{\vk' - \vk} - |\dispersion(\vk') - \dispersion(\vk)| ),
\end{equation}
where $\Theta(k)$ is the Heaviside function and $\dispersion(k) = \electronDispersion (k) + \eFock(k)$ is the single-particle dispersion.
The Fock energy $\eFock (k)$ arises from the mean-field treatment of $H_\mathrm{C}$, see \eqref{eqn:fock_coulomb} below.
We will, furthermore, restrict our discussion to a single phonon mode at the Debye frequency, i.e., $\omega_{\vk' - \vk} = \debye$,
and a constant interaction strength $G \coloneqq 2 |M_{\vk' - \vk}|^2 / \debye > 0$.
This yields the interaction term
\begin{equation}
  \label{eqn:phonon_int}
  g_{\text{CUT}}(k, k') = - \frac{G}{2} \Theta( \debye - |\dispersion(k) - \dispersion(k')| ).
\end{equation}
Note that the factor $2$ cancels with the spin summation in \eqref{eqn:phonon_hamiltonian}.
We will refer to this interaction as the \emph{\CUTinteraction{}} throughout this article.

Within the framework of this article, we will only consider this interaction in the BCS channel.
The full interaction exhibits divergent resonances at $|\omega_{\vk' - \vk}| = |\epsilon_{k'} - \epsilon_{k}|$ in the Fock channel.
These resonances cannot be avoided by other types of transformations because they appear in the renormalization of the single-particle dispersion which is an on-shell process.
Such resonances would likely be supressed by lifetime effects or smeared out when considering dispersive phonons. 
But this is beyond the scope of this article.
Thus, we restrict our discussion to the interaction \eqref{eqn:phonon_hamiltonian} as described above.

We fix the Fermi wavevector to $\hbar \kf / \sqrt{m_e} = 4.25 \sqrt{\mathrm{eV}}$ throughout this article.
This results in a Fermi energy of $\SI{8.5}{eV}$ to $\SI{9}{eV}$ depending on the screening wavevector $k_s$.
Additionally, we fix the Debye frequency to $\debye = \SI{10}{meV}$.
These values are taken to be close to the actual parameters found in lead (Pb) \cite{kittel2004}. 
We stress that our results are qualitatively independent of these parameters.

%%%%% Parameters of lead (Pb)
% E_F = 9.37 eV
% Theta_Debye = 105 K
% omega_Debye = 9.04 meV
% Delta = 2.73 meV

Furthermore, we define the dimensionless parameter 
\begin{equation}
  g \coloneqq \rho_\mathrm{F} G = G \frac{m_e \kf}{2 \pi^2 \hbar^2},
\end{equation}
where $\rho_\mathrm{F}$ is the density of states of the non-interacting system at the Fermi edge.
The derivation of the effective interaction is perturbative in the second order of the electron-phonon coupling $M$.
This perturbative approach is well justified if the matrix element $M$ is sufficiently small relative to the phonon energy $\debye$. 
We emphasize nevertheless, that large values of $g$ are possible if the density of states at the Fermi edge $\rho_F$ is large. 
Hence, it is justified to focus here on moderate values of $g$ of the order 1 to 10.

In the same manner, we define a dimensionless scaling ratio $\lambda$ for the screening wavevector
\begin{equation}
  k_s \coloneqq \lambda \sqrt{\frac{e^2 m_e}{3 \pi^2 \hbar^2 \epsilon_0} \kf}.
\end{equation}
The square-root expression is an estimate for the Thomas-Fermi wavevector in real materials \cite{rickayzen1980}.
Mathematically, this screening translates to the real-space potential $V(r) \propto \exp (- k_s r) / r$.
Physically, its existence can be motivated in the following manner:
An electric field naturally causes a displacement of the charge carriers implying a polarization which in turn modifies the effective electric field.
Diagrammatically, this process results in an effective interaction as depicted in \cref{fig:feynman},
which corresponds to the aforementioned screened Coulomb potential \cite{rickayzen1980}.

Lastly, we define the following operators for brevity
\begin{equation}
  \opf{\vk} \coloneqq \opSC{k},\qquad \opn{\vk}{\sigma} \coloneqq \ladder[\dagger]{k}{\sigma} \ladder{k}{\sigma}.
\end{equation}

%%%%%%%%%%%%%%%%%%%%%%%%%%%%%%%%%%%%%%%%%%%%%%%%%%%%%%%%%%%%%%%%%%%%%%%%%%%%%%%%%%%%%%%%%%%%%%%%%%%%%%%%%%%%%%%%%%%%%
%%%%%%%%%%%%%%%%%%%%%%%%%%%%%%%%%%%%%%%%%%%%%%%%%%%%%%%%%%%%%%%%%%%%%%%%%%%%%%%%%%%%%%%%%%%%%%%%%%%%%%%%%%%%%%%%%%%%%
%%%%%                                                 Mean-Field                                                %%%%%
%%%%%%%%%%%%%%%%%%%%%%%%%%%%%%%%%%%%%%%%%%%%%%%%%%%%%%%%%%%%%%%%%%%%%%%%%%%%%%%%%%%%%%%%%%%%%%%%%%%%%%%%%%%%%%%%%%%%%
%%%%%%%%%%%%%%%%%%%%%%%%%%%%%%%%%%%%%%%%%%%%%%%%%%%%%%%%%%%%%%%%%%%%%%%%%%%%%%%%%%%%%%%%%%%%%%%%%%%%%%%%%%%%%%%%%%%%%

\section{The superconducting gap function}
\label{sec:mean_field}

\subsection{General considerations}

We use a mean-field approximation to decouple the interaction terms in \eqref{eqn:base_hamiltonian}.
This will grant us access to the expectation values and the gap functions of interest.

On the mean-field level, \eqref{eqn:phonon_hamiltonian} boils down to a single non-vanishing term
\begin{equation}
  \tilde{H}_\mathrm{Ph} = \sum_{\vk} \deltaPh (k) \opf[\dagger]{\vk} + \mathrm{H.c.},
\end{equation}
where the phononic contribution to the superconducting gap function is given by
\begin{align}
\label{eqn:delta_phonon}
  \deltaPh (k) &= \frac{1}{\volume} \sum_{\vk'} g(k,k') \expec{\opf{\vk '}} = \frac{1}{2 \pi^2} \int_0^\infty \mathrm{d} k' k'^2 g(k, k') \expec{ \opf{k '} }.
\end{align}
Repeating the same decoupling for the Coulomb repulsion $H_\mathrm{C}$ yields
\begin{align}
  \tilde{H}_\mathrm{C} &= \sum_{\vk} \deltaC (k) \opf[\dagger]{\vk} + \mathrm{H.c.} + \sum_{\vk \sigma} (\eFock(k) + \fockCorrection(k)) \opn{\vk}{\sigma}.
\end{align}
Its additional contribution to the gap function reads
\begin{equation}
\label{eqn:delta_coulomb}
  \deltaC (k) = \frac{e^2}{8 \pi^2 \epsilon_0 k} \int_0^\infty \mathrm{d} q \expec{\opf{q}} q \ln \left( \frac{k_s^2 + (q + k)^2}{k_s^2 + (q - k)^2} \right).
\end{equation}
Note that this term has the opposite sign of the term in \eqref{eqn:delta_phonon}.

Assuming $\expec{\opn{k}{\sigma}} = \Theta (\kf - |\vk|)$ at zero temperature, the Fock energy reads
\begin{align}
\label{eqn:fock_coulomb}
  \eFock (k) = - \frac{e^2}{4 \pi^2 \epsilon_0} \kf \bigg[1 &+ \frac{k_s}{\kf } \left( \arctan \left( \frac{k - \kf}{k_s} \right) - \arctan \left( \frac{k + \kf}{k_s} \right) \right) \nonumber \\
  &+ \frac{\kf^2 - k^2 + k_s^2}{2 k \kf} \ln \left( \frac{k_s^2 + (\kf + k)^2}{k_s^2 + (\kf - k)^2} \right) \bigg].
\end{align}
In the superconducting phase, however, $\expec{\opn{k}{\sigma}}$ is more complicated than assumed above.
We define the difference between these expression as $\delta_n (k) \coloneqq \Theta (\kf - k) - \expec{\opn{k}{\sigma}}$.
Notably, this quantity is mostly close to $0$ and has a finite contribution only close to the Fermi edge.
Nevertheless, the fact that $\expec{\opn{k}{\sigma}}$ in the superconducting phase is not a Heaviside function implies a small correction to the Fock energy given by
\begin{equation}
\label{eqn:fock_corr}
  \fockCorrection(k) = \frac{e^2}{8 \pi^2 \epsilon_0 k} \int_0^\infty \mathrm{d} q \delta_n (q) q \ln \left( \frac{k_s^2 + (q + k)^2}{k_s^2 + (q - k)^2} \right),
\end{equation}
which is typically of the order of a few \textmu{}eV.

Lastly, note that the Fock energy vanishes if we restrict our calculations to the BCS channel \eqref{eqn:bcs_coulomb_hamiltonian}.
At the mean-field level, the vanishing of the Fock energy is also the main difference compared to utilizing the entire Coulomb interaction.
The most notable effect of the vanishing of the Fock energy is that the Fermi energy shifts by $\eFock(\kf) \approx -\SI{0.5}{eV}$ for a weak screening of $\lambda=10^{-4}$.
Naturally, this affects $\rho_\mathrm{F}$ and, therefore, the magnitude of the gap function.
Nevertheless, its shape remains qualitatively the same.
Accordingly, to avoid repetition, 
we will limit the discussion of the mean-field results to the full Coulomb interaction \eqref{eqn:coulomb_hamiltonian}.

\begin{figure}
  \centering
  \includegraphics[width=0.5\columnwidth]{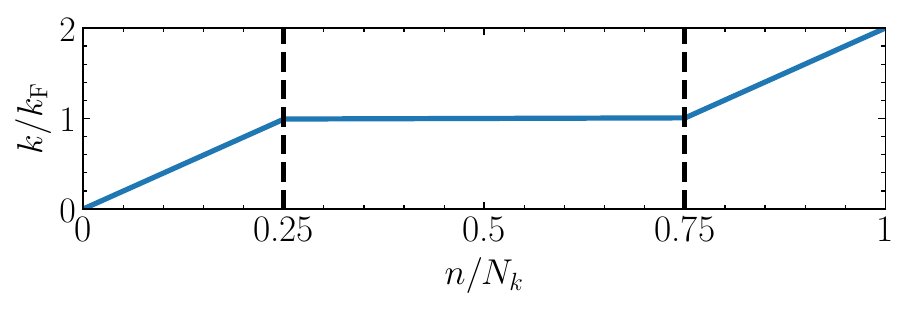}
  \caption{Schematic representation of the numerical discretization showing the magnitude of $k$ versus the discretization index $n$.
  The black dashed lines show the cutoff points at which the discretization switches from coarse to fine and vice versa.
  Note that there is still a minute slope for $n \in [N_k / 4, 3 N_k / 4]$. }
  \label{fig:discretization}
\end{figure}

For the numerics, we notice that all of the above quantities merely depend on the absolute value $k = |\vk|$ 
due to the rotational symmetry of the model.
Therefore, we discretize $k$ using $N_k$ individual points. 
Due to the long-range nature of the Coulomb interaction,
we need to extend our discretization to a large interval, theoretically even from $0$ to $\infty$.
But in practice, the choice $k \in [0, 2 \kf]$ yields good results.

The main physics can be observed around the Fermi edge so that an equidistant discretization in the entire interval would be inefficient.
It is more efficient to use a fine mesh around the Fermi edge that utilizes $N_k / 2$ individual points and is cut off at $\kf \pm x_\mathrm{cut} \debye / \kf$.
Beyond this cutoff, the discretization is coarser with $N_k / 4$ additional discretization points in each direction.
A schematic representation is shown in \cref{fig:discretization}.

The choice of $x_\mathrm{cut}$ is arbitrary to some extent.
If the gap is small, a small value is beneficial to resolve the relevant part of the gap function in more detail.
If the gap is large, however, we require a larger $x_\mathrm{cut}$ so that the entire relevant part is included.
The relevant metric is the order parameter's value at its maximum $\maxGap$.
In practice, we choose $x_\mathrm{cut} = 10$ for $\maxGap < \SI{4}{meV}$, $x_\mathrm{cut} = 20$ for $\maxGap < \SI{14}{meV}$, and $x_\mathrm{cut} = 25$ beyond that. 
These cutoffs allow for some leeway, i.e., varying these values slightly does not impact the results, as long as one chooses a large enough $N_k$.
In practice, $N_k = 30000$ produces good results for $x_\mathrm{cut}=25$, while $N_k = 20000$ is sufficiently precise for $x_\mathrm{cut} = 10$ and $20$.

\subsection{Discussion of the numerical results}

\begin{figure}
  \centering
  \includegraphics[width=0.9\columnwidth]{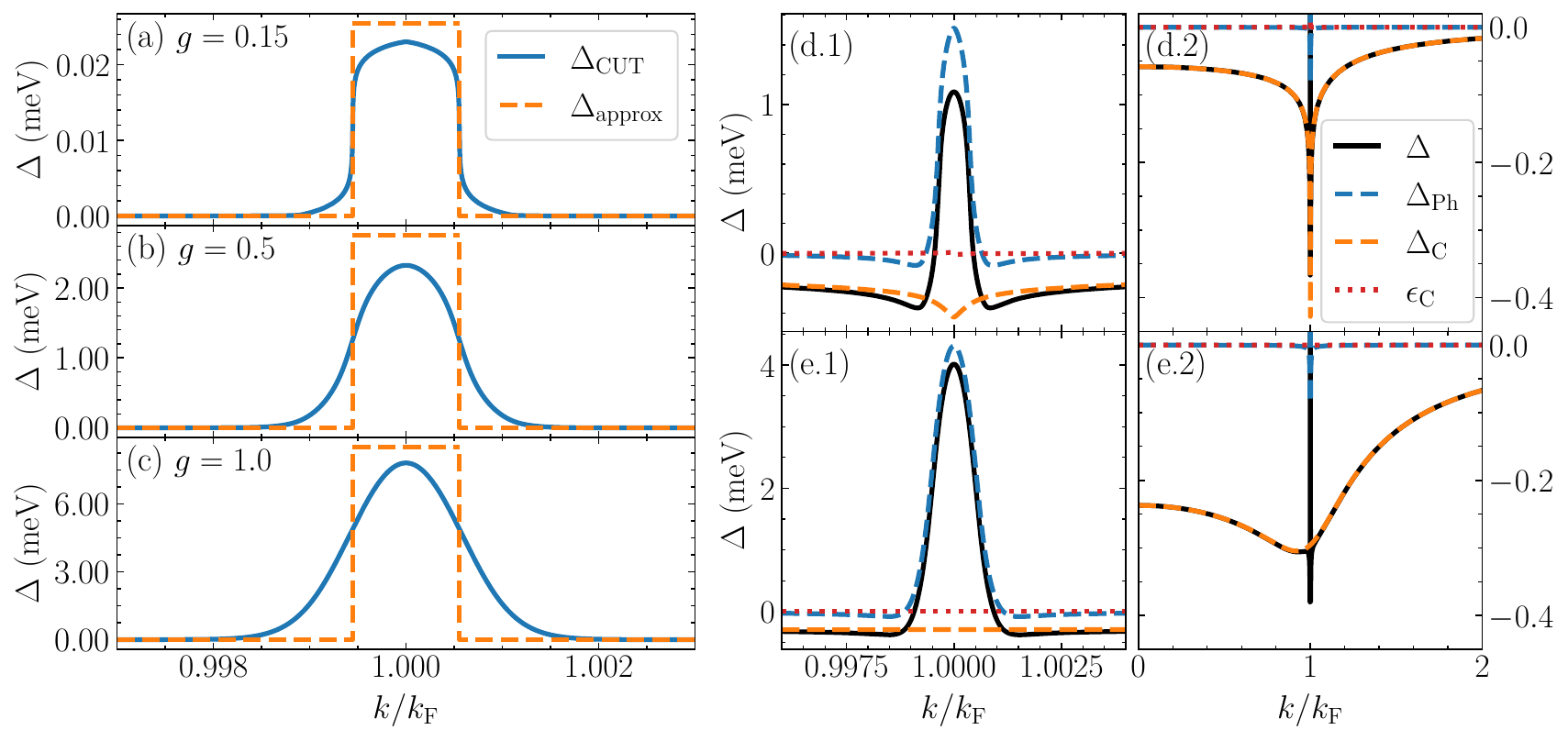}
  \caption{(a-c) The superconducting gap function without Coulomb interaction. 
  The blue solid line shows the result for the \CUTinteraction{} \eqref{eqn:phonon_int}
  while the orange dashed line depicts the data for the \BCSinteraction{} \eqref{eqn:approx_g}.
  In each case, we set $N_k = 8000$ and $g$ according to the text in each panel.
  (d-e) The superconducting gap function including the Coulomb interaction. 
  The blue and orange dashed lines depict the phononic \eqref{eqn:delta_phonon} 
  and Coulomb \eqref{eqn:delta_coulomb} contributions to the gap function,
  which itself is represented by the black solid line.
  The red dotted line shows the correction to the Fock energy \eqref{eqn:fock_corr}.
  We used $\lambda = 10^{-4}$ for panels (d) and $\lambda = 1$ for panels (e).
  Panels (1) show the gap functions in a small region around the Fermi edge,
  while panels (2) depict them across the entire numerical range.
  Note the difference in energy scales between the panels.
  For both cases, we set $N_k = 20000$ and $g=0.8$.}
  \label{fig:gap_all}
\end{figure}

Let us begin by omitting the Coulomb interaction, i.e., setting $e=0$.
In this case, the gap function only includes the phononic contribution.
Often, the effective electron-electron interaction is assumed to be constant in a small region around the Fermi edge \cite{bardeen1957,bogoliubov1958,kittel1963}.
It is then approximated by
\begin{equation}
\label{eqn:approx_g}
  g_{\text{BCS}}(k, k') \coloneqq -\frac{G}{2} \Theta(\debye - |\dispersion(k)|) \Theta(\debye - |\dispersion(k')|).
\end{equation}
We will refer to this interaction as the \emph{\BCSinteraction{}} throughout this article.
We briefly study this case as well to compare the results to those obtained by using the proper description \eqref{eqn:phonon_int}.
Three exemplary gap functions for \cref{eqn:approx_g} and the \CUTinteraction{} \eqref{eqn:phonon_int}, are depicted in \cref{fig:gap_all} (a-c).
We set $N_k = 8000$ and $g=\{0.15, 0.5, 1.0\}$. The gap functions themselves are qualitatively generic.
Increasing $g$ increases their magnitude and increasing $\debye$ increases their width.

As expected, both gap functions are restricted to a small region around the Fermi edge.
However, the gap function corresponding to the \CUTinteraction{} is slightly wider and does not have hard boundaries.
It further exhibits a smaller magnitude.
The gap function's general form approaches the form obtained via the \BCSinteraction{} in the limit $g \to 0$.

Often, the focus of studies lies on the gap's magnitude at the Fermi edge.
To this end, the approximation \eqref{eqn:approx_g} is certainly sufficient for a qualitative description of the physics, although the results differ slightly.
Nevertheless, we will see important differences between both approaches later during the discussion of the collective excitations.

Next, we add the Coulomb interaction which results in considerable changes to the gap function.
A plot of two exemplary gap functions is shown in \cref{fig:gap_all} (d) and (e).
Here, we used $N_k = 20000$ and $g=0.8$. 
In the top panel (d), the screening is set to $\lambda = 10^{-4}$ while it is set to $\lambda = 1$ in the bottom panel (e).
Again, the functions are qualitatively generic with regard to the dependence on $g$ and $\debye$.
Naturally, increasing the screening also increases the gap's magnitude, see \cref{fig:gap_all} (d.1) and (e.1).
Also, as mentioned before, the correction to the Fock energy $\fockCorrection$ is negligibly small.

A striking feature that arises due to the Coulomb interaction is that the gap function stretches over all $k$ values.
This occurs for any screening strength.
While its main contribution is still confined to a small region around the Fermi wavevector, 
it approaches a constant value for $k \to 0$ and follows a $1/k^2$ behavior as $k \to \infty$. 
This statement is proven in Appendix \ref{sec:limiting_gaps}.

Furthermore, due to the repulsive nature of the Coulomb interaction, 
the gap function has a radial node when said interaction starts dominating over the phononic one.
Similar behavior is found within the framework of the Eliashberg theory and the Anderson-Morel model \cite{carbotte1990,vashishta1973,joas2002,sigrist2005}.

At this point, it should be mentioned that the complex phase of the gap function can still be fixed to be real.
We repeated the calculation allowing for an arbitrary complex phase, but the results are essentially identical.
Instead of changing the sign at the nodal points, the complex phase of the gap function jumps by $\pi$.

Considering a small screening, i.e., $\lambda = 10^{-4}$ in panels (d) of \cref{fig:gap_all}, 
results in a rather narrow valley in $\deltaC$ around the Fermi edge.
It is noteworthy, that this result is essentially identical to the one obtained by omitting the screening entirely 
because no screening is necessary to deal with the logarithmic singularities as the latter are smoothed by the integration.
While this part of our calculations can still be easily performed without screening, the later parts require at least a small screening to avoid numerically dividing by $0$.

The aforementioned valley in $\deltaC$ disappears entirely if we use the screening wavevector introduced in Ref. \cite{rickayzen1980}, i.e., $\lambda = 1$ in panel (e.1).
In this case, the contribution of $\deltaC$ around the Fermi edge is essentially a constant.
This corroborates the qualitative statements of the commonly used result by Morel and Anderson \cite{morel1962}
that a screened Coulomb interaction essentially boils down to a constant pseudopotential $\mu$ around the Fermi edge, 
modifying the relevant interaction strength $g \to g - \mu$ \cite{mielke1997,sigrist2005,kostrzewa2018,fischer2018,simonato2023}.
Bogoliubov provided a first expression for the pseudopotential \cite{bogoliubov1958}.
We multiply it by the density of states at the Fermi edge $\rho_\mathrm{F}$ to obtain the dimensionless expression
\begin{equation}
  \label{eqn:bogoliubov_mu}
  \mu = \frac{e^2 \rho_F }{4 \epsilon_0 \kf^2 } \ln \left( \frac{k_s^2 + 4 \kf^2}{k_s^2} \right).
\end{equation}
Later, Morel and Anderson showed that, in the weak-coupling limit $\mu \ll 1$, the relevant interaction strength is actually given by $g - \mu^*$ with
\begin{equation}
  \label{eqn:anderson_morel_mu}
  \mu^* = \frac{\mu}{1 + \mu \ln \left( \frac{\Ef}{\debye} \right)},
\end{equation}
which they finally used to obtain \cite{morel1962}
\begin{equation}
  \label{eqn:anderson_morel_gap}
  \ln \left( \frac{\maxGap}{2 \debye} \right) = -\frac{1}{g - \mu^*}.
\end{equation}

\begin{figure}[!h]
  \centering
  \includegraphics[width=0.5\columnwidth]{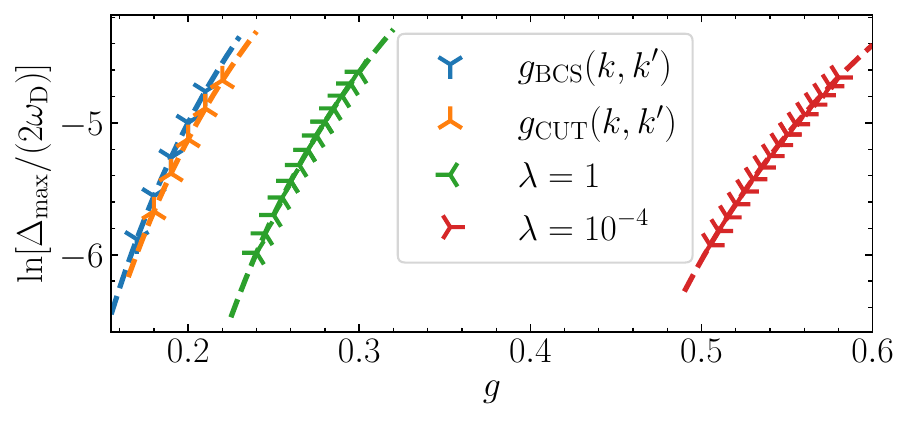}
  \caption{Logarithm of $\maxGap$ in units of $2 \debye$ versus the phononic coupling strength $g$.
  The markers represent the evaluated data points while the lines show the best fit in accordance with \cref{eqn:fit_gap}.
  The data represented by the blue and orange markers excludes the Coulomb interaction and used the \BCSinteraction{} \eqref{eqn:approx_g} and the \CUTinteraction{} \eqref{eqn:phonon_int}, respectively.
  For the data in green and red, we used the \CUTinteraction{} and included the Coulomb interaction while fixing the screening to $\lambda=1$ and $\lambda=10^{-4}$.
  The fit parameters are given in \cref{tab:fit_gap_dependence}.}
  \label{fig:fit_gap_dependence}
\end{figure}

Using this expression, we can fit our results for $\maxGap$. 
Note, however, that \cref{eqn:anderson_morel_gap} was derived under the assumption of a constant \BCSinteraction{}.
We already saw in \cref{fig:gap_all} (a-c) that the \CUTinteraction{} \eqref{eqn:phonon_int} leads to a slightly reduced gap value at the Fermi edge.
Thus, we introduce two additional fit parameters $\alpha$ and $\beta$ and fit to
\begin{equation}
  \label{eqn:fit_gap}
  \ln \left( \frac{\maxGap}{2 \debye} \right) = \ln \alpha - \frac{\beta}{g - \mu^*}.
\end{equation}
Here, $\alpha$ represents a constant ratio relating $\maxGap$ to $\debye$, while $\beta$ serves as the scaling factor in the exponential.
Based on \eqref{eqn:anderson_morel_gap}, we expect these values to be at least close to $1$.
In \cref{fig:fit_gap_dependence}, we show the fits of the data excluding and including the Coulomb interaction.
Without the Coulomb interaction, we investigated both the \BCSinteraction{} \eqref{eqn:approx_g} and the \CUTinteraction{} \eqref{eqn:phonon_int}.
Including the Coulomb repulsion, we show only the \CUTinteraction{} with the screening being set to $\lambda=1$ and $\lambda=10^{-4}$.
The corresponding fit parameters are provided in \cref{tab:fit_gap_dependence}.

\begin{table}[!h]
  \centering
  \caption{Fit parameters for the fits shown in \cref{fig:fit_gap_dependence}. 
  The data sets are obtained by 
  (i) using the \BCSinteraction{} \eqref{eqn:approx_g}, 
  (ii) using the \CUTinteraction{} \eqref{eqn:phonon_int},
  (iii) additionally including the Coulomb repulsion with $\lambda=1$, and
  (iv) using $\lambda=10^{-4}$.
  We also show the values of the pseudopotential proposed by Bogoliubov \eqref{eqn:bogoliubov_mu}, denoted as $\mu$, 
  and the pseudopotential $\mu^*$ of Morel and Anderson \eqref{eqn:anderson_morel_mu}.}
  \label{tab:fit_gap_dependence}
  \begin{tabular}{c@{\hspace{1\tabcolsep}}ccc@{\hspace{1\tabcolsep}}cc}
  \toprule
  Data  &            \multicolumn{3}{c}{Fit parameters}                             & \multicolumn{2}{c}{Predictions} \\ \cmidrule(lr) {2-4} \cmidrule(lr) {5-6}
        & $\alpha$              & $\beta$               & $\mu_{\text{Fit}}^*$      & $\mu$    & $\mu^*$  \\ \midrule
  (i)   & $1.0000 \pm 0.0003$   & $0.9999 \pm 0.0001$   & $(8 \pm 9) \cdot 10^{-6}$ & $0$      & $0$      \\
  (ii)  & $0.7804 \pm 0.0009$   & $0.9679 \pm 0.0004$   & $0.00154   \pm 0.00003$   & $0$      & $0$      \\
  (iii) & $0.707  \pm 0.012$    & $1.056  \pm 0.006$    & $0.0526    \pm 0.0005$    & $0.0688$ & $0.0469$ \\
  (iv)  & $0.31   \pm 0.04$     & $1.02   \pm 0.06$     & $0.291     \pm 0.006$     & $0.3428$ & $0.1035$ \\ \bottomrule
  \end{tabular}
\end{table}

Since the approximation \eqref{eqn:approx_g} is just the standard BCS formulation,
we expect the BCS behavior $\ln(\maxGap / (2 \debye)) = -1/g$,
which is within numerical accuracy exactly what we find.
The \CUTinteraction{} \eqref{eqn:phonon_int} deviates from that only slightly, namely in $\alpha$, i.e., the ratio between $\maxGap$ and $\debye$ which is marginally smaller now.
Notably, the exponential scaling factor $\beta$ agrees well with the expectation $\beta=1$ in all cases.

Switching on the Coulomb interaction provides more insight.
Firstly, using a typical screening $\lambda=1$ results in a $\mu^*$ that is in reasonable agreement with \cref{eqn:anderson_morel_mu}.
A small screening of $\lambda=10^{-4}$, however, yields considerable deviations.
Nevertheless, this is no contradiction to the original statement as it was derived for $\mu \ll 1$, which is no longer given here.

\begin{figure}[!t]
  \centering
  \includegraphics[width=0.5\columnwidth]{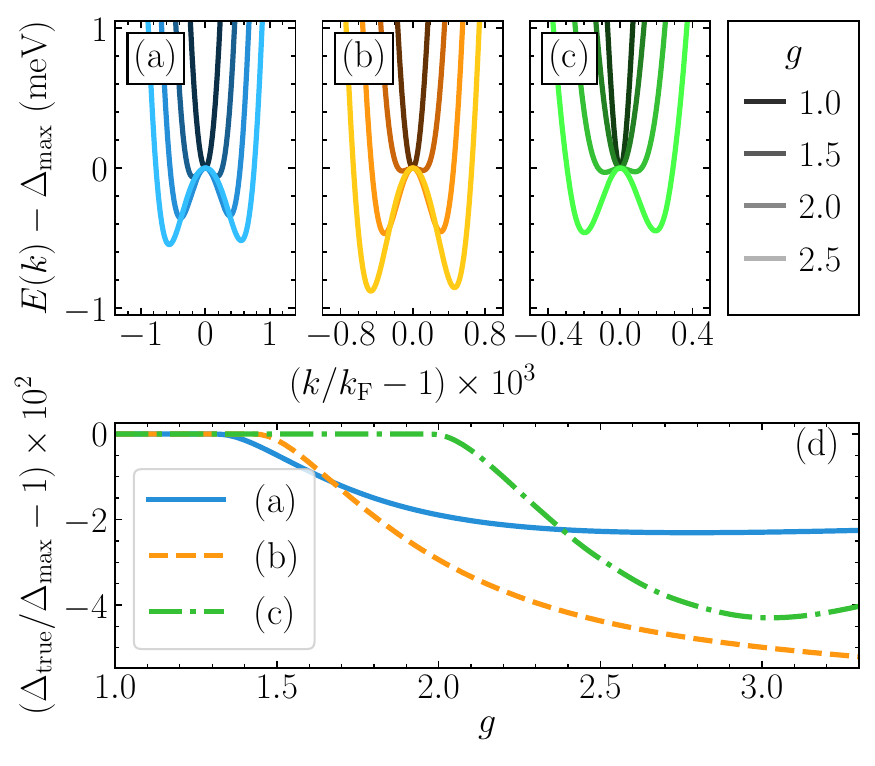}
  \caption{The upper row (a-c) depicts the quasiparticle dispersion \eqref{eqn:quasiparticle_energy} relative to the order parameter's peak value $\maxGap$.
  The individual panels have (a) no Coulomb interaction, (b) the screening set to $\lambda = 1$, and (c) $\lambda = 10^{-4}$.
  Brighter lines represent stronger phononic interactions in accordance with the legend in the top right.
  Notably, the energy minimum is not located at $\kf$ for large $g$ but shifts to $k < \kf$.
  The plot is not entirely symmetrical about $\kf$ due to the parabolic dispersion, though the effect is minuscule.
  The lower panel (d) depicts the system's true energy gap $\trueGap$ in units of order parameter's peak value $\maxGap$ versus the phononic interaction strength $g$.
  The individual lines represent the form of the Coulomb interaction as described above in accordance with the legend.
  Note that the $x$-axis begins at $g=1$ because $\trueGap = \maxGap$ holds for moderate $g$.}
  \label{fig:max_true}
\end{figure}
\begin{figure}[!ht]
  \centering
  \includegraphics[width=0.5\columnwidth]{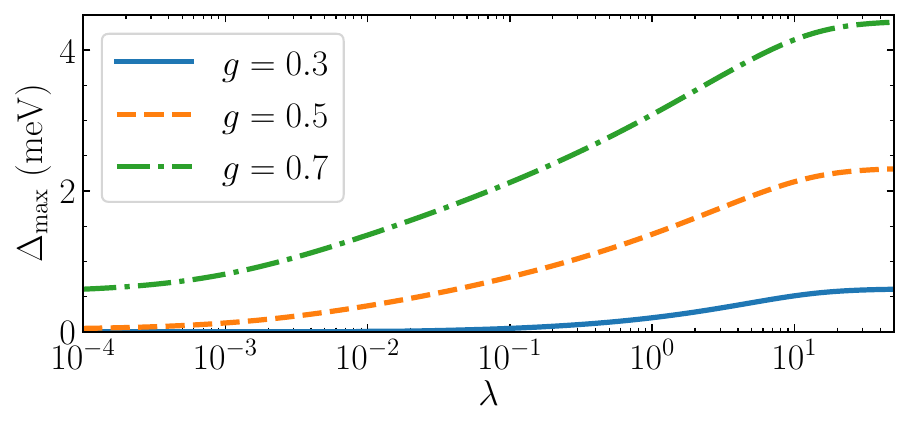}
  \caption{Plot of the order parameter's peak value $\maxGap$ as a function of the screening $\lambda$.
  The screening is scaled logarithmically. 
  The phononic interaction strength is set to three distinct values, see the legend.
  Otherwise, the parameters are identical to the ones in \cref{fig:gap_all} (d) and (e).}
  \label{fig:gap_behavior}
\end{figure}

If we assume a constant interaction strength and neglect the Coulomb interaction,
we obtain a constant order parameter $\Delta (k)$ around the Fermi edge.
Then, it is trivial to see that the quasiparticle dispersion
\begin{equation}
  \label{eqn:quasiparticle_energy}
  E(k) = \sqrt{ (\dispersion(k) + \fockCorrection(k))^2 + \Delta^2 (k) }
\end{equation}
has its global minimum at the $\kf$.
However, this is not necessarily the case if the order parameter varies.
In principle, it could diminish faster than the single-particle dispersion $\dispersion(k)$ rises for $k\neq \kf$.
We do observe this behavior for large enough $g$, see \cref{fig:max_true}. 
In the upper panels (a-c), we show the quasiparticle energy \eqref{eqn:quasiparticle_energy} relative to $\maxGap$ for various $g$.
The Coulomb interaction is excluded in panel (a) and included with a screening of $\lambda=1$ in panel (b) and $\lambda=10^{-4}$ in panel (c).
For all cases, there exists a $g$ with $\min (E(k)) \neq \kf$, though $|k_\text{min} - \kf| / \kf \ll 1$ holds.
We define the energy at this minimum as $\trueGap$, and compare it to $\maxGap$ in panel (d) of \cref{fig:max_true}.
Initially, the ratio is a constant at $\trueGap / \maxGap - 1 = 0$ since both quantities are identical for moderate $g$.
Subsequently, the minimum shifts away from $\kf$ and $\trueGap$ diminishes.
Nevertheless, this behavior slows down and even reverses as $g$ is increased further.

Lastly, \cref{fig:gap_behavior} shows the order parameter's peak value $\maxGap$ depending on the screening $\lambda$.
Naturally, it grows as the screening is increased, but it does not seem to follow any elementary function.
Expectedly, the function approaches a constant value for both cases $\lambda \to 0$ and $\lambda \to \infty$.
The general behavior does not vary qualitatively for various values of $g$.

%%%%%%%%%%%%%%%%%%%%%%%%%%%%%%%%%%%%%%%%%%%%%%%%%%%%%%%%%%%%%%%%%%%%%%%%%%%%%%%%%%%%%%%%%%%%%%%%%%%%%%%%%%%%%%%%%%%%%
%%%%%%%%%%%%%%%%%%%%%%%%%%%%%%%%%%%%%%%%%%%%%%%%%%%%%%%%%%%%%%%%%%%%%%%%%%%%%%%%%%%%%%%%%%%%%%%%%%%%%%%%%%%%%%%%%%%%%
%%%%%                                               Mode analysis                                               %%%%%
%%%%%%%%%%%%%%%%%%%%%%%%%%%%%%%%%%%%%%%%%%%%%%%%%%%%%%%%%%%%%%%%%%%%%%%%%%%%%%%%%%%%%%%%%%%%%%%%%%%%%%%%%%%%%%%%%%%%%
%%%%%%%%%%%%%%%%%%%%%%%%%%%%%%%%%%%%%%%%%%%%%%%%%%%%%%%%%%%%%%%%%%%%%%%%%%%%%%%%%%%%%%%%%%%%%%%%%%%%%%%%%%%%%%%%%%%%%

\section{Collective excitations}
\label{sec:mode_analysis}

\subsection{General considerations}

Analyzing collective excitations is more involved and lies beyond the simple mean-field approach discussed in the previous section.
We follow the iEoM approach as discussed in Ref. \cite{althuser2024} to obtain the relevant Green's functions.

To investigate Higgs and phase modes, we compute the Green's function with respect to the operators
\begin{subequations}
  \begin{align}
      \label{eqn:higgs_operator}
      \mathfrak{A}_\mathrm{Higgs} &\coloneqq \frac{1}{\sqrt{\volume}} \sum_{\vk} \left( \opf[\dagger]{\vk} + \opf{\vk} \right), \\
      \label{eqn:phase_operator}
      \mathfrak{A}_\mathrm{Phase} &\coloneqq \frac{\im}{\sqrt{\volume}} \sum_{\vk} \left( \opf[\dagger]{\vk} - \opf{\vk} \right).
  \end{align}
\end{subequations}
\edited{Note that both operators are fully isotropic, i.e., any modes excited by them are modes without angular momentum.
Furthermore, these operators have no center-of-mass momentum. 
Consequently, the corresponding excited collective excitations are excitations at zero momentum.
To compute the corresponding Green's functions}, we choose an operator basis $\mathfrak{B}$ for the iEoM that contains the operators
\begin{subequations}
\begin{align}
  \mathcal{A}_k &\coloneqq \frac{1}{\sqrt{\volume}} \sum_{\vq} \delta (k - |\vq|) \left( \opf[\dagger]{\vq} + \opf{\vq} \right) \\
  \mathcal{P}_k &\coloneqq \frac{1}{\sqrt{\volume}} \sum_{\vq} \delta (k - |\vq|) \left( \opf[\dagger]{\vq} - \opf{\vq} \right) \\
  \mathcal{N}_k &\coloneqq \frac{1}{\sqrt{\volume}} \sum_{\vq} \delta (k - |\vq|) \left( \opn{\vq}{\up} + \opn{-\vq}{\down} \right)
\end{align}
\end{subequations}
for all $k$ under consideration.
Note that \eqref{eqn:higgs_operator} and \eqref{eqn:phase_operator} are easily represented by our choice of basis.
For example, consider
\begin{align}
  \mathfrak{A}_\mathrm{Phase} &= \frac{\im}{\sqrt{\volume}} \int \mathrm{d} k \sum_{\vq} \delta(k - |\vq|) 
                                      \left( \opf[\dagger]{\vq} - \opf{\vq} \right) = \im \int \mathrm{d} k \mathcal{P}_k.
\end{align}
The next step is to compute the dynamical matrix $\mathcal{M}$ and norm matrix $\mathcal{N}$
\bs
\begin{align}
  \mathcal{M}_{ij} &= (O_i | [H, O_j]) \\
  \mathcal{N}_{ij} &= (O_i | O_j)
\end{align}
\es
with each $O_i \in \mathfrak{B}$ and the pseudo-scalar product $(A | B) \coloneqq \expec{ [A^\dagger, B] }$.
We evaluate the occurring commutators with respect to the full Hamiltonian \eqref{eqn:base_hamiltonian}.
Then, we take the expectation values with respect to the mean-field Hamiltonian
and make use of Wick's theorem to compute quartic expectation values.

These matrices can be analytically related to the Fourier-transformed Green's functions
\begin{equation}
\label{eqn:green_function_definition}
  \greenSymbol_{\alpha} (z = \omega + \im 0^+) =
     -\frac{\im}{\hbar} \int_0^\infty \mathrm{d}t e^{\im zt} \expec{ [\mathfrak{A}_{\alpha} (t), \mathfrak{A}_{\alpha}^\dagger (0)] },\quad 
     \alpha \in \{ \mathrm{Higgs}, \mathrm{Phase} \}.
\end{equation}
Numerically, we perform a Lanczos tridiagonalization and obtain a continued fraction expansion of \cref{eqn:green_function_definition} that we terminate using the square-root terminator \cite{althuser2024,pettifor1984,viswanath1994}.
Then, the spectral functions are obtained by $\spectralSymbol_{\alpha} = - (1/\pi) \Im [\greenSymbol_{\alpha} (\omega + \im 0^+)]$.

As the relevant physics occurs in the vicinity of the Fermi edge, we restrict the analysis of the collective modes to this region.
Specifically, the $k$-integrals in this section are restricted to the fine mesh around the Fermi edge.
This procedure allows us to obtain precise data for the collective modes that have an energy similar to or smaller than the gap function.
However, it carries the caveat that the two-particle continuum is only resolved up to energies in the order of $x_\mathrm{cut} \debye$.
One can expect that the same applies to other features at high energies, i.e., 
that it will not be possible to provide precise numerical data for collective excitations at high energies.
Still, we will see for example that the phase mode shifts to large energies as we include the Coulomb repulsion.

Lastly, it should be mentioned, that we require at least some screening for these calculations because the commutators yield otherwise divergent terms.
As an example consider 
$\expec{[ \opf[\dagger]{\vk'} - \opf{\vk'} , [H,  \opf{\vk} - \opf[\dagger]{\vk} ]]}$.
One of the resulting terms is $V( |\vk - \vk'| )  \langle \opf{ \vk } \rangle \langle \opf{ \vk' } \rangle$,
which has a singularity at $\vk = \vk'$.
Because these terms are contributions to individual matrix elements of matrices that we need to manipulate in a non-trivial manner,
we cannot lift nor easily avoid these kinds of singularities and ultimately require a finite screening.
But we find that the results are not very sensitive to the exact value of $\lambda \ll 1$.

\subsection{Classification of collective excitations}

\begin{figure}
  \centering
  \includegraphics[width=0.5\columnwidth]{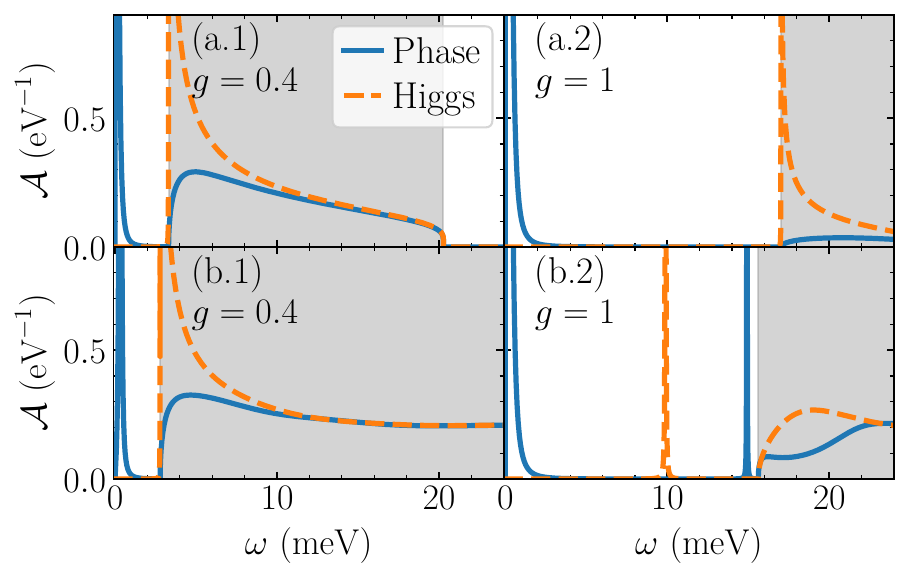}
  \caption{Spectral functions $\spectral{Phase}$ (blue solid line) and $\spectral{Higgs}$ (orange dashed line) without the Coulomb interaction.
  A small imaginary shift is introduced to the energy $\omega \to \omega + \im \delta$ to plot the $\delta$-peaks below the continuum as Lorentzians with $\delta = 10^{-3}\,\mathrm{meV}$.
  In the upper panels (a), we used the \BCSinteraction{} \eqref{eqn:approx_g} while we used the \CUTinteraction{} \eqref{eqn:phonon_int} in the lower ones.
  The grey area marks the two-particle continuum.
  Prominently, we can identify the phase mode at $\omega=0$, and, for small $g$ the Higgs mode at $\omega=2\maxGap$.
  The Higgs mode moves below the gap and additional modes appear as $g$ is increased, but only for the \CUTinteraction{} \eqref{eqn:phonon_int}.}
  \label{fig:resolvent_overview}
\end{figure}

As before, we start the discussion by omitting the Coulomb interaction, i.e., $e=0$.
First, we compute the spectral functions using the \BCSinteraction{} \eqref{eqn:approx_g}.
The results are shown in the upper panels of \cref{fig:resolvent_overview}. 
Here, we shift the energy into the complex plane by a small constant $\omega \to \omega + \im \delta$ to resolve the peaks below the two-particle continuum.
In this article, we set $\delta = 10^{-3}\,\mathrm{meV}$.
The interaction strength was set to $g=0.4$ in the left panels and to $g=1$ in the right panels,
which yields a gap of $\maxGap \approx \SI{1.65}{meV}$ and $\maxGap \approx \SI{8.51}{meV}$, respectively.
The spectral functions each exhibit a single peak, $\spectral{Higgs}$ at $\omega = 2 \maxGap$ and $\spectral{Phase}$ at $\omega = 0$.
These features are well-known in the literature and are commonly associated to the Higgs and the phase mode, respectively 
\cite{schmid1975,varma2002,krull2016,schwarz2020,cea2014,measson2014,tsuji2015,reinhoffer2022,sulaiman2024,fischer2018,althuser2024,anderson1958,bogoliubov1958,brieskorn1974,simanek1975,schon1976,kulik1981,maiti2015,sun2020,fan2022,volkov1973,yuzbashyan2006,dzero2024}.

To analyze peaks below the continuum, we fit $\Re [\greens{\alpha}]$ in close vicinity to them.
For the phase peak at $\omega=0$, we find $\Re [\greens{Phase}] \propto 1/\omega^2$,
which translates to $\spectral{Phase} \propto \delta'(\omega)$ due to the Kramers-Kronig relations.
This can be interpreted as two peaks at $\pm \omega_0$ with $\omega_0 \to 0$ 
because the spectral function with respect to a Hermitian bosonic operator is antisymmetric \cite{althuser2024}.

By fitting $\spectralSymbol_\mathrm{Higgs}(\omega)$ close to the peak, we find that it falls off like $1 / \sqrt{\omega - 2 \maxGap}$.
These results are consistent with previous findings.
For instance, the order parameter behaves as $\Delta(t) \propto \cos(2\Delta t) / \sqrt{t}$ shortly after a weak interaction quench \cite{kulik1981,volkov1973,yuzbashyan2006,dzero2024} which is the Fourier transform of our result. 
Also note that there is no qualitative difference between the results obtained for different values of $g$.

Next, we examine the \CUTinteraction{} \eqref{eqn:phonon_int}.
The results are displayed in the lower panels of \cref{fig:resolvent_overview}.
For small to moderate interaction strengths $g$, the results do not differ significantly from those obtained using the \BCSinteraction{}.
The Higgs mode continues to behave asymptotically like an inverse square root while the phase mode behaves like $\delta'(\omega)$.
Notably, the most significant difference is that $\maxGap$ is slightly reduced.
Specifically, the gaps are now $\maxGap \approx \SI{1.39}{meV}$ and $\maxGap \approx \SI{7.81}{meV}$
instead of $\maxGap \approx \SI{1.65}{meV}$ and $\maxGap \approx \SI{8.51}{meV}$, respectively.

For larger values of $g$, however, the Higgs mode shifts to energies below $2 \maxGap$.
Increasing the interaction strength further spawns even more modes below the continuum in both $\spectral{Higgs}$ and $\spectral{Phase}$, see \cref{fig:resolvent_overview} panel (b.2).
Note that this only happens if we use the \CUTinteraction{} \eqref{eqn:phonon_int} and not if we use the \BCSinteraction{} \eqref{eqn:approx_g}.
This indicates that there are intricate mechanisms at play that are not captured by the simple approximation \eqref{eqn:approx_g}.

To distinguish between the modes, we refer to the Higgs mode and the phase mode that can be associated with those found in literature as \emph{primary} modes.
The additional modes that emerge below the continuum are referred to as \emph{secondary} modes.

\begin{figure}[!t]
  \centering
  \includegraphics[width=0.5\columnwidth]{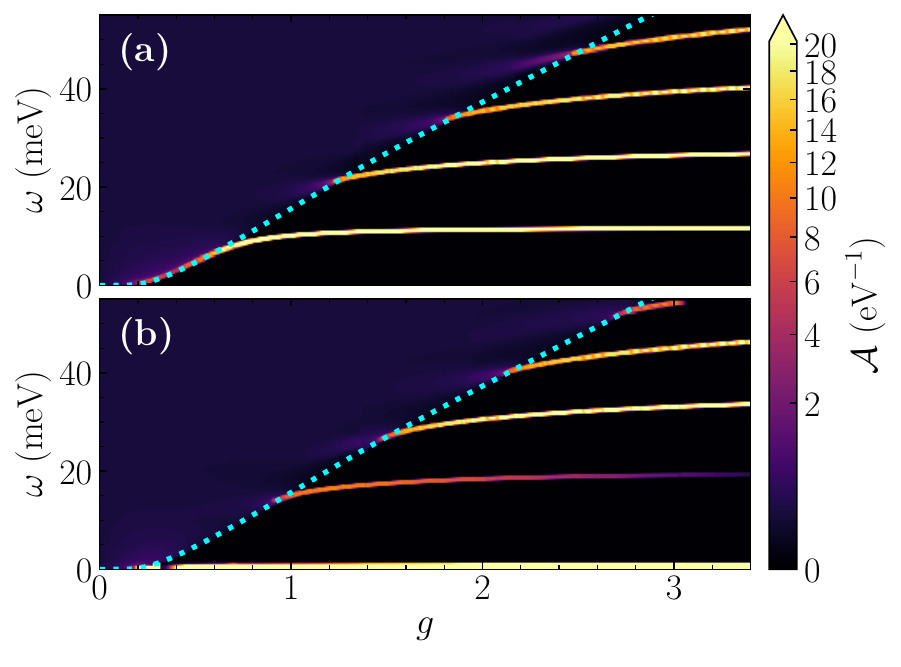}
  \caption{Spectral functions (a) $\spectral{Higgs}$ and (b) $\spectral{Phase}$ without the Coulomb interaction, i.e., $e=0$.
  The magnitude is represented by the color scale. 
  The peaks below the continuum are mathematically $\delta$-peaks and are represented by Gaussian bell curves with the same weight and width $\sigma=\SI{0.05}{meV}$.
  The only exception to this is the peak at $\omega=0$ in $\spectral{Phase}$.
  It is described by the derivative of a $\delta$-peak and therefore represented by the derivative of a Gaussian bell curve.
  The coupling strength $g$ is indicated on the $x$-axis.
  The cyan dotted line shows the lower edge of the two-particle continuum at $2\trueGap$.
  The evolution of the modes is clearly visible as they emerge smoothly from the two-particle continuum.
  Note that the color scale is non-linear to better depict the occurring modes.}
  \label{fig:heatmap_full_nc}
\end{figure}

As anticipated, the primary phase mode remains located at $\omega = 0$, 
since there are still no long-range Coulomb interactions present \cite{goldstone1961,anderson1963}.
Furthermore, we fit the real part of the Green's functions near the peaks below the continuum.
Doing so yields the same behavior as before for the primary phase peak 
and $\Re [\greens{Phase}] \propto 1/(\omega - \omega_0)$ for the peaks at finite energies $\omega_0 < 2 \maxGap$
indicating $\spectral{\alpha} \propto \delta(\omega - \omega_0)$.
In the same manner, it is straightforward to determine the spectral weights of these excitations.

To observe the evolution of the collective excitations upon varying $g$, 
we plot the spectral functions $\spectralSymbol (\omega)$ in \cref{fig:heatmap_full_nc}.
We build on the previous analysis and represent the occurring $\delta$-peaks by Gaussian bell curves and the $\delta'$-peaks by the derivative of a Gaussian bell.
The energy $\omega$ is plotted on the $y$-axis and the phononic coupling strength on the $x$-axis.
The coloring indicates the magnitude of the spectral functions.
The modes appear as sharp bright lines, while the two-particle continuum is faintly visible due to the comparatively low magnitude of the spectral functions inside that region.

To guide the eye, we also plot the lower edge of the continuum at $2 \trueGap$ as a cyan dotted line.
Notably, the primary phase mode remains at $\omega=0$ for all $g$ as expected.

The primary Higgs mode first evolves along the lower edge of the continuum until it smoothly breaks away from it for larger $g$.
\edited{This phenomenon has been observed previously by Barankov and Levitov in the context of order parameter oscillations after an interaction quench \cite{barankov2007}.
Note that the non-trivial momentum dependence of the interaction is essential to observing this detachment of the Higgs mode.
For the constant \BCSinteraction{}, the Higgs mode remains at $2\maxGap$.}
After that, the mode has a minute upward tendency in energy. 
At the same time, additional modes emerge smoothly from the continuum.
These \bonusModes{} appear in an alternating fashion first in the phase and then in the Higgs spectral function.
Their energy gain remains small afterward.
We will further investigate these modes in \cref{sec:bonus_modes}.

\subsection{Effects of the Coulomb interaction}

\begin{figure}[!t]
  \centering
  \includegraphics[width=0.5\columnwidth]{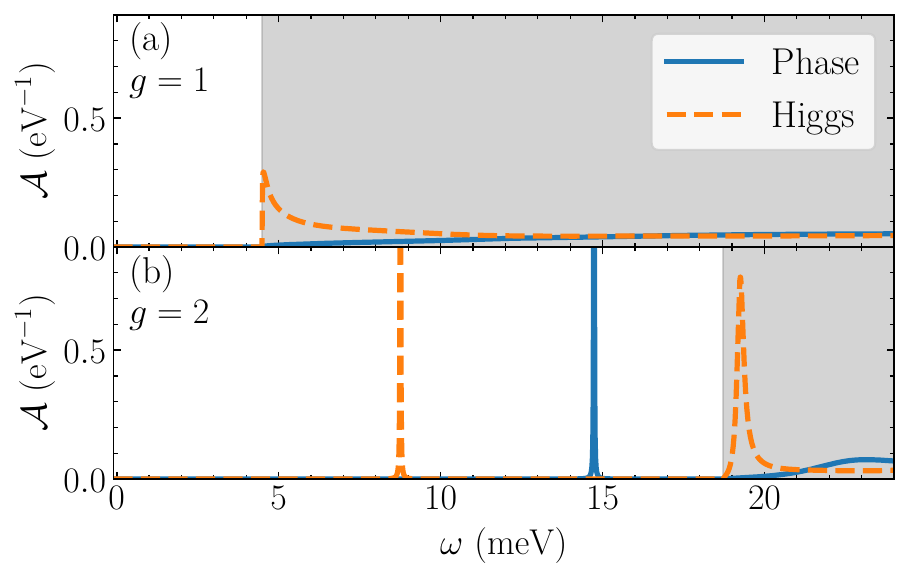}
  \caption{Spectral functions $\spectral{Phase}$ (blue solid line) and $\spectral{Higgs}$ (orange dashed line) with the Coulomb interaction.
  A small imaginary shift is introduced to the energy $\omega \to \omega + \im \delta$ to resolve the peaks below the continuum; $\delta = 10^{-3}\,\mathrm{meV}$.
  Panel (a) and (b) depict the results for $g=1$ and $g=2$, respectively. 
  For both cases, we used an almost vanishing screening $\lambda=10^{-4}$.
  The Higgs mode persists after the introduction of the Coulomb interaction.
  The phase mode at $\omega=0$, however, disappears as expected and shifts to high energies beyond the regime depicted in this plot. 
  Increasing $g$ again spawns \bonusModes{}, similar to what is presented in \cref{fig:resolvent_overview}.
  }
  \label{fig:resolvent_overview_coulomb}
\end{figure}

To establish a baseline for the following discussions, we include the full Coulomb interaction with a nearly vanishing screening $\lambda = 10^{-4}$.
The resulting spectral functions are depicted in \cref{fig:resolvent_overview_coulomb}.
Setting $g=1$, results in $\maxGap \approx \SI{2.2}{meV}$.
The Higgs mode is diminished in magnitude, but its asymptotic behavior remains $\spectral{Higgs} \propto 1 / \sqrt{\omega - 2 \maxGap}$.
The phase mode, however, is absent from the low-energy regime and can now be found at high energies, as one would expect.

Increasing $g$ spawns the same \bonusModes{} as before, but this effect only manifests at significantly larger values of $g$ compared to before.
This shift can be attributed to the Coulomb repulsion, which reduces the gap. 
Consequently, a stronger attraction $g$ is required to achieve the same $\maxGap$ as before. 
Therefore we postulate that the governing mechanism behind the occurrence of these excitations is not the attraction strength $g$ itself, but the magnitude of the gap.
We will delve into this statement later, cf. \cref{sec:bonus_modes}.

Note that there is still no mode at $\omega=0$ in \cref{fig:resolvent_overview_coulomb} panel (b).
Additionally, the lowest-lying Higgs mode appears at lower energies than the lowest-lying phase mode.
With this in mind, we want to investigate how the phase mode shifts to higher energies.
(i) Is the BCS channel of the interaction sufficient to explain this behavior?
(ii) Does it evolve smoothly in terms of the screening?

\begin{figure}
  \centering
  \includegraphics[width=0.9\columnwidth]{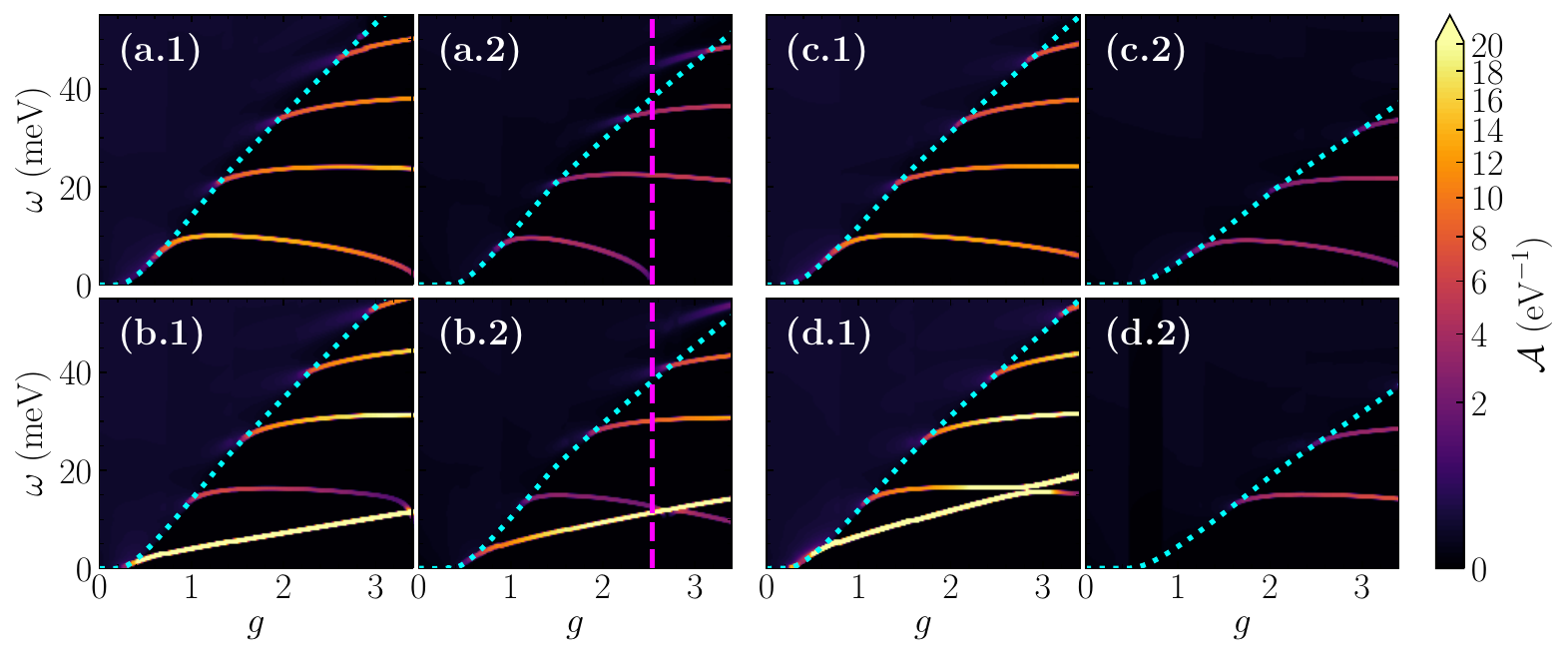}
  \caption{Same as \cref{fig:heatmap_full_nc}, except that the Coulomb interaction is included.
  We used only in the BCS channel \eqref{eqn:bcs_coulomb_hamiltonian} for panels (a) and (b), 
  while we used the full Coulomb interaction \eqref{eqn:coulomb_hamiltonian} for panels (c) and (d).
  The upper panels (a) and (c) depict $\spectral{Higgs}$ and the lower panels (b) and (d) depict $\spectral{Phase}$.
  Panels (1) show the data for a screening of $\lambda = 1$ and panels (2) the data for $\lambda = 10^{-4}$.
  Notably, the phase mode is no longer located at $\omega=0$ but at finite energy below the two-particle continuum for all cases except panel (d.2).
  Additionally, when the lowest-lying Higgs mode in panel (a.2) reaches $0$ (magenta dashed line at $g \approx 2.54$), 
  the dynamical matrix $\mathcal{M}$ picks up a negative eigenvalue, hinting at an instability of the system.}
  \label{fig:heatmap_full_big}
\end{figure}

To address question (i), we use the Coulomb interaction only in the BCS channel \eqref{eqn:bcs_coulomb_hamiltonian}
and perform the computations for various $g$.
We show the resulting data in \cref{fig:heatmap_full_big} (a) and (b).
As before, the upper panels (a) depict $\spectral{Higgs}$ and the lower panels (b) depict $\spectral{Phase}$.
In the left column (1), we used a screening of $\lambda = 1$. For the right one (2), we set $\lambda=10^{-4}$.

The results for both cases are qualitatively similar.
The \bonusModes{} emerge from the continuum essentially as they did without any Coulomb interactions.
The primary Higgs mode, however, exhibits an altered behavior.
It picks up a downward tendency for both screenings and even reaches $\omega=0$.
For $\lambda=10^{-4}$, this point is $g \approx 2.54$, which is marked by a magenta dashed line, see \cref{fig:heatmap_full_big} (a.2).
This behavior indicates instabilities of the system.
Simultaneously, the dynamical matrix $\mathcal{M}$ acquires a negative eigenvalue for $g > 2.54$.
However, this matrix must be positive semidefinite if the system is in thermal equilibrium \cite{althuser2024}.
Thus, we conclude that beyond that point the system favors a different phase, which is beyond the scope of this article, though.

Turning to $\spectral{Phase}$ in the lower panels, we notice that the primary phase mode is present below the continuum.
For both screenings, it shifts to higher energies linearly with $g$.
At a certain point, it crosses the next-higher phase mode, which exhibits a downward tendency just as the Higgs mode did.
Such level crossings do not occur typically, as couplings between the two levels introduce level repulsion.
Combined with the fact that the primary phase mode remains at relatively small energies,
we conclude that the BCS channel for the Coulomb interaction is insufficient to fully describe the relevant physics.
Nevertheless, including the Coulomb interaction only in the BCS channel suffices to lift the primary phase mode to a finite energy.

Hence, we turn to the study of the full Coulomb interaction \eqref{eqn:coulomb_hamiltonian} in \cref{fig:heatmap_full_big} (c) and (d).
First considering $\lambda=1$ in the left panels (1), we find that the modes are mostly similar to the ones presented earlier.
However, the level crossing is now absent and has been replaced by an anticrossing.

The energies of the lowest Higgs and the lowest phase mode still have a downward tendency at larger $g$ for both screening strengths.
Beyond the region shown in the plot for $g>3.65$, the same behavior as before occurs, i.e., the primary Higgs mode shifts to $\omega=0$ and the system becomes unstable.
We will discuss this in more detail later, cf. \cref{sec:bonus_modes} and \cref{fig:mode_weights} (a.3).

\begin{figure}[!t]
  \centering
  \includegraphics[width=0.9\columnwidth]{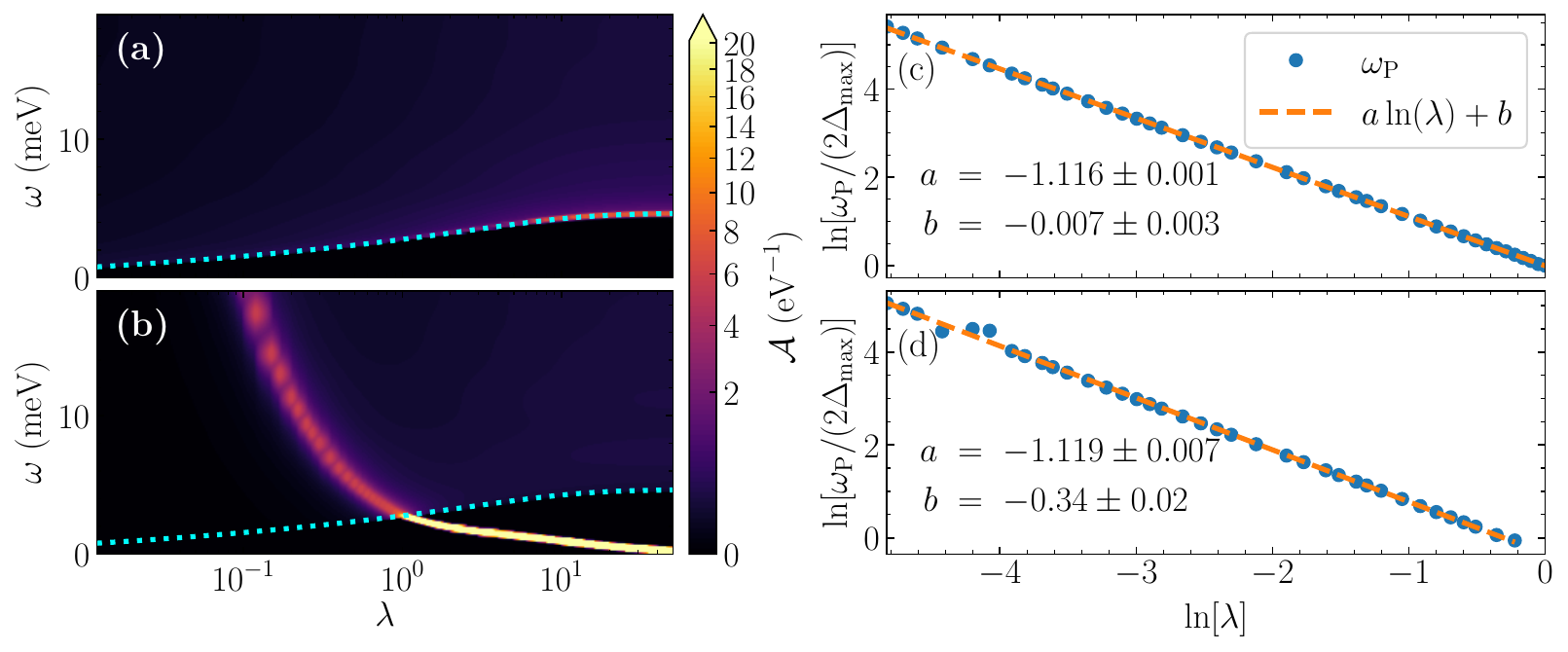}
  \caption{The left panels depict the spectral functions (a) $\spectral{Higgs}$ and (b) $\spectral{Phase}$ with the full Coulomb interaction at different screenings $\lambda$.
  The phononic coupling strength is fixed to $g=0.5$ and $\lambda$ is varied logarithmically on the $x$-axis.
  The magnitude is represented by the color scale. 
  The color scale is non-linear to depict the occurring modes better. 
  The peaks below the continuum are mathematically $\delta$-peaks and are represented by Gaussian bell curves with the same weight and width $\sigma=\SI{0.05}{meV}$.
  The cyan dotted line shows the lower edge of the two-particle continuum at $2\trueGap = 2\maxGap$.
  The phase mode smoothly shifts down from high energies as the screening is increased.
  The Higgs mode is located at $\omega=2\maxGap$ for all screenings because the parameters do not yield large gaps.
  The right panels show double logarithmic plots of the position $\omega_\mathrm{P}$ of the resonance in $\spectral{Phase}$ in units of $2\maxGap$ depending on the screening $\lambda$.
  The phononic interaction strength is set to (c) $g=0.5$ and (d) $g=0.7$.
  The blue circles mark the data used for the fits which are represented by the orange dashed lines.
  The plot depicts a range of $\lambda$ so that $\omega_\mathrm{P}$ is inside the continuum.
  Here, $\omega_\mathrm{P}$ follows an almost straight line with a slope independent of $g$.}
  \label{fig:heatmap_lambda}
\end{figure}

A central point is that we confirm that the primary phase mode vanishes for all $g$ from the low-energy regime.
Thereby, we are able to provide a numerical calculation that is fully in unison with the Anderson-Higgs mechanism \cite{anderson1958,schon1976,kulik1981}.
Moreover, the BCS channel merely couples the Cooper pairs to one another while the full interaction accounts for the interaction between all electrons.
Combining this with the fact that the BCS channel is insufficient to describe the Anderson-Higgs mechanism,
we can confirm that the phase mode is coupled to the collective motion of the electrons by the inclusion of the Coulomb interaction. 
This coupling occurs specifically due to the inclusion of a nearly unscreened interaction with proper long-range behavior.

The next goal is to understand how the panels (c.1) and (d.1) of \cref{fig:heatmap_full_big} evolve into the panels (c.2) and (d.2) upon decreasing $\lambda$, thereby answering question (ii).
To this end, we fix the phononic interaction strength to $g=0.5$ and vary the screening $\lambda$.
The results are shown in \cref{fig:heatmap_lambda} (a) and (b) where the $\lambda$-axis is scaled logarithmically.
As before, the upper panel (a) depicts $\spectral{Higgs}$, and the lower one (b) depicts $\spectral{Phase}$.
Of course, the gap is affected by the screening since a stronger screening implies weaker repulsion and therefore a larger gap.
Other than that, there is little impact on the Higgs mode.
It remains located at $\omega=2\maxGap$ and it only gains in magnitude as the gap itself grows.
Note that the Higgs mode does not detach from the two-particle continuum because the parameters do not result in large gaps.
In contrast, tracing the phase mode from vanishing screening toward strong screening, we observe its smooth evolution down from high energies.
It leaves the continuum around $\lambda=1$, though the precise value depends on the specific choice of system parameters.

To better understand this process, we show the phase mode's position $\omega_\mathrm{P}$ in units of $2 \maxGap$ versus the screening in a double-logarithmic plot in \cref{fig:heatmap_lambda} (c) and (d).
We depict the result for (c) $g=0.5$ and (d) $g=0.7$. 
The plot is restricted to the screenings for which $\omega_\mathrm{P}$ is inside the continuum because the fits only work in this range.
Within the continuum, $\omega_\mathrm{P}$ behaves almost like $1/\lambda$, possibly with logarithmic corrections, as indicated by the fits represented by orange dashed lines.

Contrary to expectations, this behavior does not stop at the plasma frequency, which in our case is a few electronvolts.
Rather, $\omega_\mathrm{P}$ continues to increase in the same manner as described above as $\lambda$ decreases.
We believe that this is due to numerical limitations.
We cannot resolve the high-energy regime properly as our approach, including the choice of the numerical mesh and cutoff, is focused on the low-energy regime.
Therefore, the mode likely behaves as depicted in \cref{fig:heatmap_lambda} for moderate screenings but will deviate in the limit $\lambda \to 0$ in accordance with literature predictions \cite{anderson1958,schon1976,kulik1981}.

Lastly, it bears mentioning that tuning $\debye$ has no significant effect on the results.
We present the data for this statement in Appendix \ref{sec:debye_frequency}.

\subsection{Emergence of \bonusModes{}}
\label{sec:bonus_modes}

\begin{figure}[!t]
  \centering
  \includegraphics[width=0.9\textwidth]{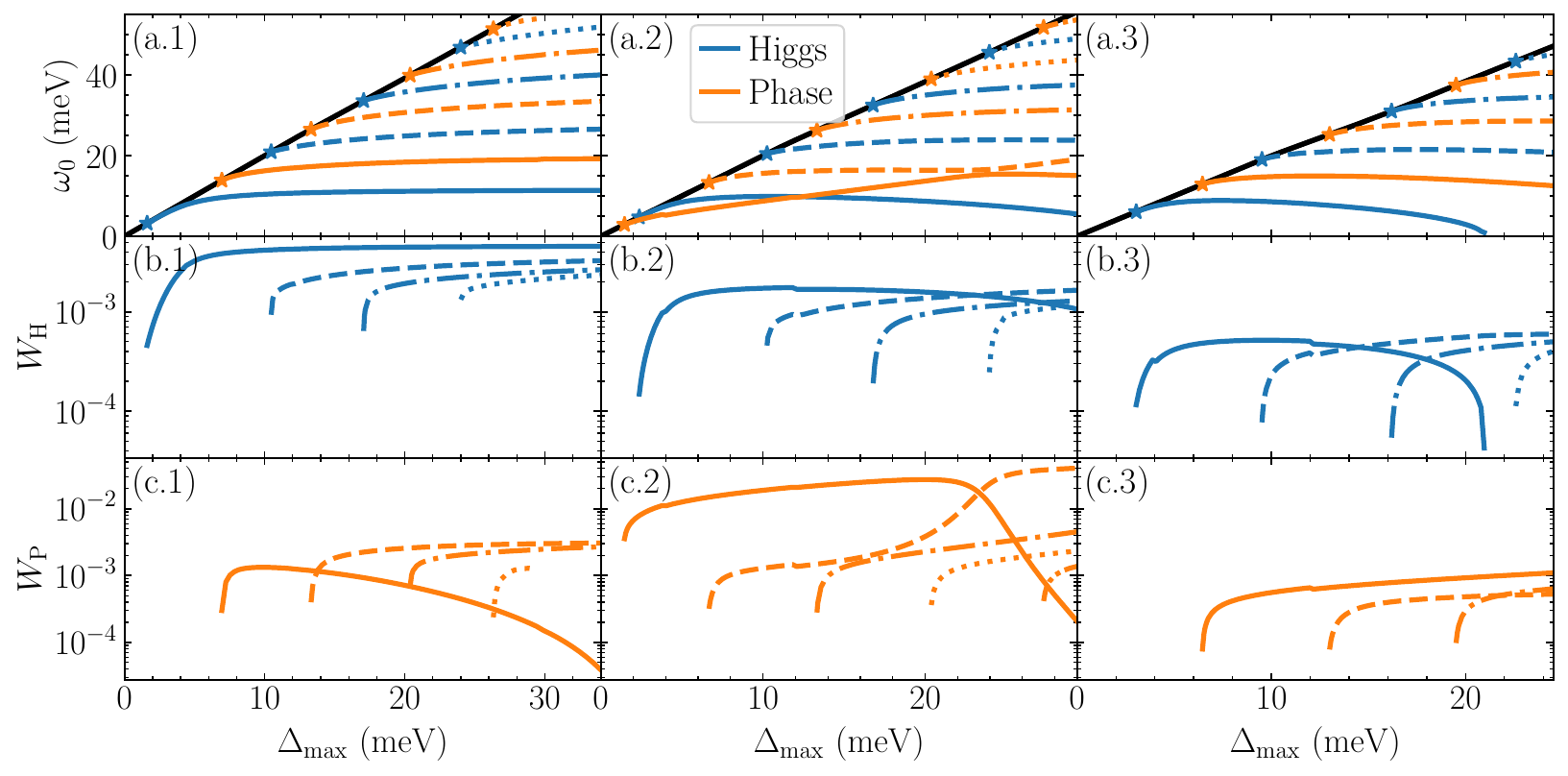}
  \caption{The panels in the first row (a) show the energies $\omega_0$ of the Higgs (blue) and phase (orange) modes depending on $\maxGap$.
  The black line marks the lower edge of the two-particle continuum. 
  In column (1) the Coulomb interaction is omitted while it is included in the other two columns; in (2) with $\lambda=1$ and in (3) with $\lambda=10^{-4}$.
  The $y$-axes of the last two rows are scaled logarithmically and depict the weight of the peaks in (b) $\spectral{Higgs}$ and (c) $\spectral{Phase}$.
  Lines with different styles correspond to one another within a column, e.g., the solid blue lines in (a.1) and in (b.1) depict the data for the same mode.
  For the left panels (1), we omit the primary phase mode at $\omega=0$ due to its distinct behavior as discussed in the previous section.
  The peaks emerge with initially vanishing weights.}
  \label{fig:mode_weights}
\end{figure}

The last question we seek to address is how the various modes behave as they emerge from the two-particle continuum.
These \bonusModes{} emerge both with and without Coulomb interactions but at different $g$.
As formulated previously, the working hypothesis is that the modes emerge at certain threshold values of the gap $\maxGap$,
not at specific values of $g$.
If the Coulomb repulsion is stronger, i.e., less screened, a larger value of $g$ is requried to induce the same gap $\maxGap$.

\Cref{fig:mode_weights} depicts the peak positions $\omega_0$ (panels (a)) and their weights $W$ (panels (b) for $\spectral{Higgs}$ and (c) for $\spectral{Phase}$) as a function of the maximum of the gap.
The left panels (1) depict the data for no Coulomb interaction.
The remaining panels show the data for the full Coulomb interaction with $\lambda=1$ in the center panels (2) and $\lambda=10^{-4}$ in the right panels (3).
We choose different linestyles to distinguish between the various modes.
The blue lines depict the modes in $\spectral{Higgs}$ and the orange lines those in $\spectral{Phase}$.
Additionally, we omitted the primary phase mode at $\omega=0$ from the left panels (1) because it has distinct properties from the secondary ones.

At the point of emergence, the peaks have vanishing weights; the mode's weight remains within the continuum.
However, as the modes move away from the continuum, the weights of the peaks increase rapidly.
Most of the excitations' weights essentially saturate beyond which they exhibit only a slight upward tendency.

One exception is the first phase mode in panel (c.1) as its weight tends towards $0$.
A similar behavior occurs in panel (c.2).
Here, the first and the second phase mode experience a level repulsion for large $g$.
At this point, their weights swap, i.e., the lower-lying mode loses most of its weight which then tends to $0$.
Concurrently, the other mode gains a significant weight and then behaves like the other modes,
i.e., its weight has only a slight upward tendency.

\begin{figure}[!t]
  \centering
  \includegraphics[width=0.5\columnwidth]{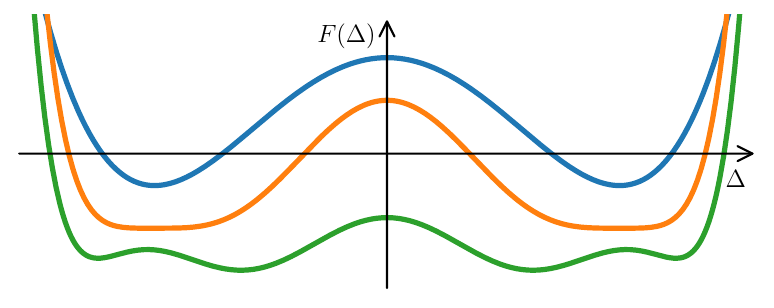}
  \caption{Schematic representation of the free energy $F$ as a function of the superconducting order parameter $\Delta$.
  The standard Landau theory results in the blue curve.
  The orange curve represents the situation when the primary amplitude mode becomes soft.
  Here, changing the order parameter barely affects the free energy.
  The green curve depicts a possible landscape of the free energy afterward when multiple minima might be present.}
  \label{fig:landau}
\end{figure}

The first Higgs mode tends towards $\omega=0$ and loses its weight for both cases that include Coulomb interactions (panels 2 and 3).
As mentioned in the previous section, this kind of behavior is a precursor to a phase transition.
When dealing with superconducting systems, one commonly employs the Landau theory for continuous phase transitions.
In this case, the free energy $F$ of the system is approximated by a polynomial in the order parameter, see \cref{fig:landau}.

The standard Landau theory results in parabolic minima at the equilibrium value of $\Delta$ as sketched by the blue curve \cite{coleman2015}.
However, small perturbations to the magnitude of the order parameter would barely affect the free energy if the free-energy landscape becomes flat around the minima as depicted by the orange curve.
This is likely what happens as the primary Higgs mode becomes soft and therefore small amplitude fluctuations require minimal energy to be excited.
Beyond this point, the free energy might exhibit multiple minima, as sketched by the green curve.

\begin{figure}[!t]
  \centering
  \includegraphics[width=0.5\columnwidth]{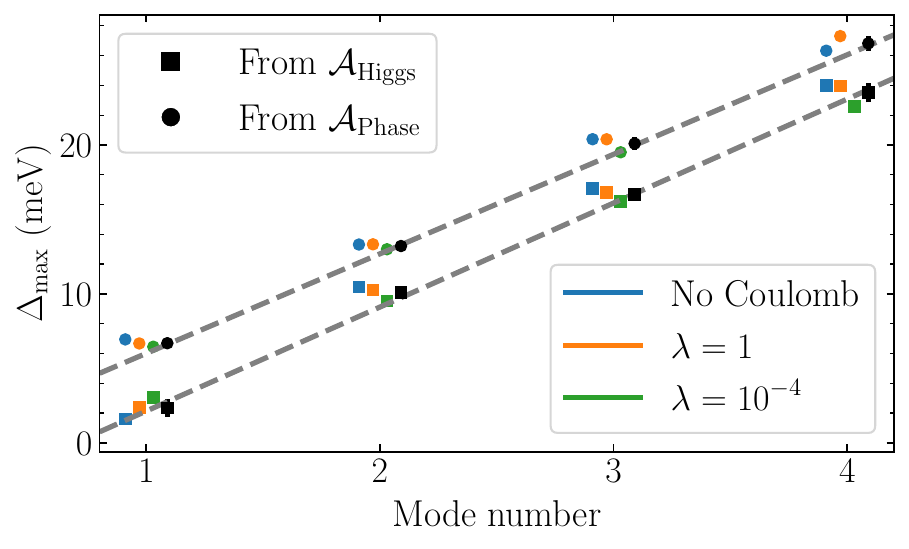}
  \caption{Peak value of the gap $\maxGap$ at which the \bonusModes{} emerge from the continuum.
  The squares represent Higgs modes, while the circles represent phase modes. 
  The primary phase modes have been omitted from this plot as they have distinctive behavior.
  The $x$-axis denotes the mode number. 
  The data points are shifted slightly from integer values to improve visibility.
  The colors indicate the type of Coulomb interaction as described by the legend.
  The black symbols are the averages of the corresponding three colored ones.
  Their error bars represent the standard deviation.
  The gray dashed lines are linear fits to the black markers. 
  Their slopes $s_{\alpha}$ describe the spacing in $\maxGap$ between the emergence of individual modes.
  The slopes are $s_{\mathrm{Higgs}} = {(6.98 \pm 0.26)}\,\mathrm{meV}$ and $s_{\mathrm{Phase}} = {(6.68 \pm 0.15)}\,\mathrm{meV}$. }
  \label{fig:mode_emergence}
\end{figure}

Upon closer inspection of the upper panels (a) in \cref{fig:mode_weights}, 
we anticipate that the various modes appear at about the same $\maxGap$, essentially independent of the Coulomb interaction.
To confirm this, we depict specifically these values in \cref{fig:mode_emergence}.
Here, the $\maxGap$ at which the modes emerge is plotted along the $y$-axis, and the modes are counted along the $x$-axis.
The square markers represent Higgs modes while the circles represent phase modes.
The colors represent the three cases of the Coulomb interaction as discussed above. 
The black markers are their averages.
The error bars represent one standard deviation.
We omitted the primary phase modes due to their special behavior.
The obvious conclusion is that corresponding excitations emerge at almost the same gaps.

We fit the averages linearly so that the slopes $s_{\alpha}$ represent the spacing in $\maxGap$ between the emergence of the modes.
This yields $s_{\mathrm{Higgs}} = {(6.98 \pm 0.26)}\,\mathrm{meV}$ and $s_{\mathrm{Phase}} = {(6.68 \pm 0.15)}\,\mathrm{meV}$, respectively.
By taking the average of these two values using inverse-variance weighting, we obtain $s = {(6.76 \pm 0.13)}\,\mathrm{meV}$.
This quantity means that for every $s$ that the gap grows, we expect to find one additional mode per channel.

Notably, repeating these calculations for $\hbar \kf / (2m) = 5\,\sqrt{\mathrm{eV}}$ yields almost the same plot.
In this case, we obtain the slopes 
$s_{\mathrm{Higgs}} = (7.04 \pm 0.25)\,\mathrm{meV}$, $s_{\mathrm{Phase}} = (6.62 \pm 0.16)\,\mathrm{meV}$, and $s = (6.75 \pm 0.14)\,\mathrm{meV}$.
This corroborates our hypothesis that the governing parameter behind the appearance of the \bonusModes{} is the magnitude of the superconducting gap.

To our knowledge, these kinds of subgap excitations have not been comprehensively studied before.
Similar excitations, however, can be found in $d$-wave superconductors.
In this case, they can be attributed to an additional rotational degree of freedom due to the nodal structure of such systems \cite{hirashima1987,schwarz2020}.

Contrary to this, our model does not break rotational symmetry.
Nevertheless, we observe a non-trivial \emph{radial} energy landscape for large phononic interaction strengths.
We presume that the emerging \bonusModes{} relate to this non-trivial radial dependence.

Specifically, the second Higgs mode emerges from the continuum around the same $g$ at which the quasiparticle dispersion forms minima at $k \neq \kf$ as discussed in \cref{sec:mean_field}, cf. \cref{fig:max_true}.
Nevertheless, this particular kind of peculiarity can be observed only for the second Higgs mode.

%%%%%%%%%%%%%%%%%%%%%%%%%%%%%%%%%%%%%%%%%%%%%%%%%%%%%%%%%%%%%%%%%%%%%%%%%%%%%%%%%%%%%%%%%%%%%%%%%%%%%%%%%%%%%%%%%%%%%
%%%%%%%%%%%%%%%%%%%%%%%%%%%%%%%%%%%%%%%%%%%%%%%%%%%%%%%%%%%%%%%%%%%%%%%%%%%%%%%%%%%%%%%%%%%%%%%%%%%%%%%%%%%%%%%%%%%%%
%%%%%                                                 Conclusion                                                %%%%%
%%%%%%%%%%%%%%%%%%%%%%%%%%%%%%%%%%%%%%%%%%%%%%%%%%%%%%%%%%%%%%%%%%%%%%%%%%%%%%%%%%%%%%%%%%%%%%%%%%%%%%%%%%%%%%%%%%%%%
%%%%%%%%%%%%%%%%%%%%%%%%%%%%%%%%%%%%%%%%%%%%%%%%%%%%%%%%%%%%%%%%%%%%%%%%%%%%%%%%%%%%%%%%%%%%%%%%%%%%%%%%%%%%%%%%%%%%%

\section{Conclusion}\label{sec:conclusion}

The goal of this article was to investigate collective excitations in superconductors using a more complete description than the standard BCS theory.
To this end, we expanded upon the standard constant attraction by employing an energy-transfer-dependent interaction \eqref{eqn:phonon_int} in the BCS channel derived via a continuous unitary transformation.
This attractive interaction is called effective phononic interaction since it stems from the electron-phonon coupling.
Additionally, we included the long-range Coulomb interaction which causes the phase mode to shift toward high energies, in accordance with the Anderson-Higgs mechanism.

We started by studying the static mean-field properties of the system.
Omitting the Coulomb interaction, the superconducting gap function behaves qualitatively similar to the one obtained by the BCS theory.
The gap is only finite in the close vicinity of the Fermi edge. Beyond this region, it tends to zero continuously.
Furthermore, the gap exhibits a slightly diminished magnitude.
Upon switching on Coulomb interactions, the gap function exhibits radial nodes close to the Fermi edge where it switches its sign as function of momentum.
We observed good agreement with the result based on the pseudopotential $\mu^*$ introduced by Morel and Anderson \cite{morel1962} if the interaction is screened.
Notable deviations occur for nearly unscreened interactions.
Here, we obtained a significantly larger value for $\mu^*$.
However, this is no contradiction because the derivation of Morel and Anderson is valid for the weak-coupling regime,
which is not applicable in the case of small screenings.

Due to the non-trivial momentum dependence of the gap function,
we found that the quasiparticle dispersion does not display its minimum at the Fermi edge for large interaction strengths.
Instead, the minimum shifts to slightly lower momenta because the gap function falls off faster than the electron dispersion rises.

The heart of the present endeavor was the study of collective excitations enabled by computing Green's functions via the iterated equations of motion approach.
By revisiting the standard BCS theory without including the Coulomb interaction, we successfully reproduced established results from the literature.
Specifically, we identified the Higgs mode at the lower edge of the two-particle continuum and confirmed the phase mode at zero energy, validating our computational framework.
Next, we went beyond the standard BCS coupling by employing the energy-transfer-dependent interaction \eqref{eqn:phonon_int} also for studying the collective excitations.

Our findings align with the previous results for small and moderate coupling strengths.
For larger coupling strengths, the Higgs mode detaches from the continuum.
\edited{Such a detachment implies that the Higgs mode becomes dissipationless as has been discussed by Barankov and Levitov in the context of persisting oscillations of the order parameter after an interaction quench \cite{barankov2007}.
The momentum dependence of the interaction is crucial for the emergence of this phenomenon.
Our results perfectly align with this as we did not see such a detachment when we considered the constant \BCSinteraction{}.
Upon increasing the coupling strength further,} additional modes emerge in both the amplitude and phase spectral functions.
To our knowledge, there has not been a comprehensive study of these isotropic subgap excitations previously.
Within the context of $d$-wave superconductors, similar excitations were found and attributed to the nodal structure of the gap function \cite{hirashima1987,schwarz2020}.
In our case, we believe that the non-trivial radial energy landscape for large phononic interaction strengths is linked to the additional, secondary modes.

\edited{Similar additional modes appear if one considers non-isotropic excitations.
In particular, Bardasis and Schrieffer used the conservation of angular momentum $L$ in isotropic systems and showed that collective excitations corresponding to different $L$ can occur below $2\Delta$ \cite{bardasis1961}.
These so-called Bardasis-Schrieffer modes are typically observed if there are strong subdominant pairing channels in the system that correspond to a different pairing symmetry such as $d$-wave superconductivity \cite{bohm2014,sun2020a,muller2021,hackner2023}.
While the ideas are similar, our results are distinct because we compute the Green's functions with respect to isotropic operators.
Therefore, the modes we identify are excitations with $L=0$.
In analogy to the standard hydrogen problem, our \bonusModes{} likely correspond to different radial dependences analogous to different values of the principal quantum number for hydrogen.}

To study the effects of the Coulomb interaction itself, we followed two approaches.
First, we restricted the Coulomb interaction to the BCS channel \eqref{eqn:bcs_coulomb_hamiltonian}.
In this case, there is little difference between the results for various screenings beyond changes of the magnitude of the gap.
This approach is sufficient to lift the phase mode from $0$ to a finite energy, however it remains below the gap.
Contrary to that, the Anderson-Higgs mechanism states that the phase mode should be located at high energies around the plasma frequency \cite{anderson1958,schon1976,kulik1981}.
Furthermore, the primary phase mode crosses with the next-higher phase mode which is at odds with the expectations for level repulsion.
Altogether, we conclude that the inclusion of the Coulomb interaction only in the BCS channel is insufficient to describe the relevant physics.

Therefore, we considered the full Coulomb interaction \eqref{eqn:coulomb_hamiltonian} next.
Crucially, we find that the primary phase mode shifts towards large energies continuously as the screening strength is decreased in line with theoretical predictions.
Our method is desigend for the low-energy regime and does not yield precise numerical values at large energies towards the upper band edge, i.e., in the limit of vanishing screening.
Nevertheless, it provides strong support for established theories of the Anderson-Higgs mechanism \cite{anderson1958,schon1976,kulik1981}.

For moderately strong screenings, we observe that the phase mode remains below the gap.
Importantly, the previously noted level crossing becomes an anticrossing as expected from level repulsion.
Thus, we confirm that it is important to include the full Coulomb interaction in all channels because it couples the phase of the order parameter to electronic density fluctuations which are not captured by the BCS channel alone.

We emphasize that the additional secondary modes persist even for the full inclusion of the Coulomb interaction.
But they appear for larger values of the coupling $g$.
Their points of emergence appear universal when they are scanned as function of the maximal value of the gap function $\maxGap$ instead of as function of the coupling $g$.
Then, these \bonusModes{} appear at regular intervals of $\maxGap$,
independent of whether Coulomb interactions are included or not.
This observation suggests that these modes represent a robust feature and will also occur in more elaborate treatments and in presence of further interactions.

Lastly, we observed the primary Higgs mode becoming soft which we interprete as the system becoming unstable at large interaction strengths.
At present, we cannot  conclude whether this feature is spurious or whether it indicates a quantum phase transition by condensation of the primary Higgs mode to novel phases.

%%%%%%%%%%%%%%%%%%%%%%%
%%%%    Outlook    %%%%
%%%%%%%%%%%%%%%%%%%%%%%

The present study calls for further investigations.
A particularly intriguing investigation in this respect is an experimental one: 
Are there systems with strong BCS coupling in which experimental signatures of the predicted secondary modes can be discerned? 
Terahertz spectroscopy appears to be an appropriate tool for this search.

On the theoretical side, at zero total momentum, the present study can be extended by allowing non-isotropic excitations in the isotropic system,
i.e., with finite angular momentum which represent Bardasis-Schrieffer modes \cite{bardasis1961,bohm2014,sun2020a,muller2021,hackner2023}.
It is conceivable that such modes become soft even before the secondary isotropic phase and Higgs modes emerge or become soft.

Another extension suggesting itself is to consider excitations not only at zero momentum,
but at finite momentum \cite{fischer2018}.
In this way, the dispersive behavior can be addressed.
But it must be pointed out that this requires more comprehensive discretizations than at zero momentum where it was sufficient to discretize the magnitude of the momentum.

To corroborate our finding of \bonusModes{} alternative approximations can be employed,
for instance a diagrammatic approach to the electron-phonon system without unitary reduction to an electron-only model.
This would amount up to analyzing the corresponding Eliashberg theory.
Thus, a future study could compare the results obtained here with those from the Eliashberg theory \cite{eliashberg1960,combescot1990,joas2002,chang2007} to understand the differences and similarities between the two approaches.
Such a comparison could not only validate our findings, but also reveal if and how lifetime effects influence the excitation spectra.

Last but not least, one can pass from the fully isotropic model to lattice models with reduced point group symmetries of the Fermi surfaces.
Then, the collective excitations above a ground state of different symmetry, such as $d$-wave symmetry \cite{schwarz2020}, can be investigated.

\section*{Acknowledgements}
We acknowledge very helpful discussions with D. Hering, J. Stolze, I. Eremin, J. Schmalian, and M. Sigrist.

% TODO: include funding information
\paragraph{Funding information}
This research was partially funded by the MERCUR Kooperation in project KO-2021-0027.

\begin{appendix}
\numberwithin{equation}{section}

\section{Limiting behavior of the gap function}
\label{sec:limiting_gaps}
In this section, we analytically show how the gap function behaves in the limits $k \to 0$ and $k \to \infty$.
First, we note that the phononic contribution \eqref{eqn:phonon_int} vanishes in both limits.
Thus, we will only discuss the Coulomb contribution \eqref{eqn:delta_coulomb}.
For $k \to 0$, consider
\begin{subequations}
\begin{align}
  \lim_{k \to 0} \deltaC (k) 
    &= \lim_{k \to 0} \frac{e^2}{8 \pi^2 \epsilon_0 k} \int_0^\infty \mathrm{d} q \expec{f_{q}^\dagger} q \ln \left( \frac{k_s^2 + (q + k)^2}{k_s^2 + (q - k)^2} \right) \\
    &= \frac{e^2}{8 \pi^2 \epsilon_0} \lim_{k \to 0} \frac{1}{k} \int_0^\infty \mathrm{d} q \expec{f_{q}^\dagger} q \left( \frac{4 k q}{k_s^2 + q^2} + \mathcal{O}(k^3) \right) \\
    &= \frac{e^2}{2 \pi^2 \epsilon_0} \int_0^\infty \mathrm{d} q \expec{f_{q}^\dagger} \frac{q^2}{k_s^2 + q^2} \\
    &= \mathrm{const}.
\end{align}
\end{subequations}
Similarly, we obtain for $k \to \infty$
\begin{subequations}
\begin{align}
  \deltaC (k \gg \kf ) &= \frac{e^2}{8 \pi^2 \epsilon_0 k} \int_0^\infty \mathrm{d} q \expec{f_{q}^\dagger} q \left( \frac{4q}{k} + \mathcal{O}\left( \frac{1}{k^3} \right) \right) \\
  &= \frac{e^2}{2 \pi^2 \epsilon_0 } \frac{1}{k^2} \int_0^\infty \mathrm{d} q q^2 \expec{f_{q}^\dagger} \\
  &\propto \frac{1}{k^2}.
\end{align}
\end{subequations}

\section{Dependence on the Debye frequency}
\label{sec:debye_frequency}

\begin{figure}
  \centering
  \includegraphics[width=0.5\columnwidth]{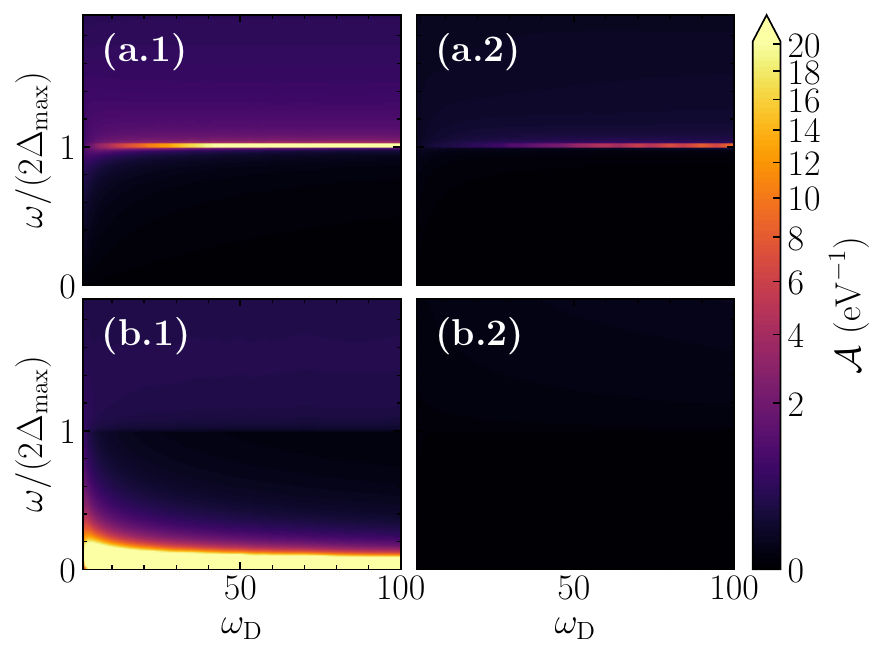}
  \caption{Spectral functions (a) $\spectral{Higgs}$ and (b) $\spectral{Phase}$ depending on $\debye$.
  The $y$-axis is scaled by $\omega / \maxGap$ to show the modes' behavior more clearly.
  The left column shows the data if no Coulomb interaction is present, while the right one includes it.
  The phononic coupling strength is set to $g=0.4$ and $g=1$, respectively.}
  \label{fig:heatmap_omega}
\end{figure}

In this section, we briefly discuss how the Debye frequency $\debye$ affects the collective excitation.
As an example, we vary $\debye$ in \cref{fig:heatmap_omega}.
For the left column (1), we omitted the Coulomb interaction and set $g=0.4$, while for the right column (2), we included it with $\lambda=10^{-4}$ and $g=1$.
The gap scales linearly with $\debye$, as discussed before.
Therefore, we plot the $\omega / (2\maxGap)$ on the $y$-axis.
The continuum begins then exactly at $1$ for moderate gap sizes.
All modes are at a constant position in this plot.
We checked this for various other parameters and found that this behavior is generic.

In conclusion, changing the Debye frequency has no significant effect on the results presented within this article.

\end{appendix}

%%%%%%%%% END TODO: CONTENTS

%%%%%%%%%% TODO: BIBLIOGRAPHY
\bibliography{main}

\begin{thebibliography}{10}
\providecommand{\url}[1]{\texttt{#1}}
\providecommand{\urlprefix}{URL }
\expandafter\ifx\csname urlstyle\endcsname\relax
  \providecommand{\doi}[1]{doi:\discretionary{}{}{}#1}\else
  \providecommand{\doi}{doi:\discretionary{}{}{}\begingroup
  \urlstyle{rm}\Url}\fi
\providecommand{\eprint}[2][]{\url{#2}}

\bibitem{bardeen1957}
J.~Bardeen, L.~N. Cooper and J.~R. Schrieffer,
\newblock \emph{Theory of {{Superconductivity}}},
\newblock Physical Review \textbf{108}(5), 1175 (1957),
\newblock \doi{10.1103/PhysRev.108.1175}.

\bibitem{bogoliubov1958}
N.~Bogoliubov,
\newblock \emph{A {{New Method}} in the {{Theory}} of {{Superconductivity}} 1},
\newblock Soviet physics JETP-USSR \textbf{7}(1), 41 (1958).

\bibitem{kulik1981}
I.~O. Kulik, O.~{Entin-Wohlman} and R.~Orbach,
\newblock \emph{Pair susceptibility and mode propagation in superconductors:
  {{A}} microscopic approach},
\newblock Journal of Low Temperature Physics \textbf{43}(5), 591 (1981),
\newblock \doi{10.1007/BF00115617}.

\bibitem{volkov1973}
A.~Volkov and {\relax Sh.M}.~Kogan,
\newblock \emph{Collisionless relaxation of the energy gap in superconductors},
\newblock Soviet physics JETP-USSR \textbf{65}(5), 2038 (1973).

\bibitem{schmid1975}
A.~Schmid and G.~Sch{\"o}n,
\newblock \emph{Linearized kinetic equations and relaxation processes of a
  superconductor near {$T_c$}},
\newblock Journal of Low Temperature Physics \textbf{20}(1), 207 (1975),
\newblock \doi{10.1007/BF00115264}.

\bibitem{varma2002}
C.~M. Varma,
\newblock \emph{Higgs {{Boson}} in {{Superconductors}}},
\newblock Journal of Low Temperature Physics \textbf{126}(3), 901 (2002),
\newblock \doi{10.1023/A:1013890507658}.

\bibitem{yuzbashyan2006}
E.~A. Yuzbashyan, O.~Tsyplyatyev and B.~L. Altshuler,
\newblock \emph{Relaxation and {{Persistent Oscillations}} of the {{Order
  Parameter}} in {{Fermionic Condensates}}},
\newblock Physical Review Letters \textbf{96}(9), 097005 (2006),
\newblock \doi{10.1103/PhysRevLett.96.097005}.

\bibitem{cea2014}
T.~Cea and L.~Benfatto,
\newblock \emph{Nature and {{Raman}} signatures of the {{Higgs}} amplitude mode
  in the coexisting superconducting and charge-density-wave state},
\newblock Physical Review B \textbf{90}(22), 224515 (2014),
\newblock \doi{10.1103/PhysRevB.90.224515}.

\bibitem{measson2014}
M.-A. M{\'e}asson, Y.~Gallais, M.~Cazayous, B.~Clair, P.~Rodi{\`e}re, L.~Cario
  and A.~Sacuto,
\newblock \emph{Amplitude {{Higgs}} mode in the
  {$2H\ensuremath{-}{\text{NbSe}}_{2}$} superconductor},
\newblock Physical Review B \textbf{89}(6), 060503 (2014),
\newblock \doi{10.1103/PhysRevB.89.060503}.

\bibitem{tsuji2015}
N.~Tsuji and H.~Aoki,
\newblock \emph{Theory of {{Anderson}} pseudospin resonance with {{Higgs}} mode
  in superconductors},
\newblock Physical Review B \textbf{92}(6), 064508 (2015),
\newblock \doi{10.1103/PhysRevB.92.064508}.

\bibitem{krull2016}
H.~Krull, N.~Bittner, G.~S. Uhrig, D.~Manske and A.~P. Schnyder,
\newblock \emph{Coupling of {{Higgs}} and {{Leggett}} modes in non-equilibrium
  superconductors},
\newblock Nature Communications \textbf{7}(1), 11921 (2016),
\newblock \doi{10.1038/ncomms11921}.

\bibitem{reinhoffer2022}
C.~Reinhoffer, P.~Pilch, A.~Reinold, P.~Derendorf, S.~Kovalev, J.-C. Deinert,
  I.~Ilyakov, A.~Ponomaryov, M.~Chen, T.-Q. Xu, Y.~Wang, Z.-Z. Gan
  \emph{et~al.},
\newblock \emph{High-order nonlinear terahertz probing of the two-band
  superconductor {${\mathrm{MgB}}_{2}$}: {{Third-}} and fifth-order harmonic
  generation},
\newblock Physical Review B \textbf{106}(21), 214514 (2022),
\newblock \doi{10.1103/PhysRevB.106.214514}.

\bibitem{sulaiman2024}
V.~Sulaiman and G.~S. Uhrig,
\newblock \emph{Optical pumping of {{Bardeen-Cooper-Schrieffer}}
  superconductors},
\newblock Physical Review B \textbf{110}(18), 184513 (2024),
\newblock \doi{10.1103/PhysRevB.110.184513}.

\bibitem{fischer2018}
S.~Fischer, M.~Hecker, M.~Hoyer and J.~Schmalian,
\newblock \emph{Short-distance breakdown of the {{Higgs}} mechanism and the
  robustness of the {{BCS}} theory for charged superconductors},
\newblock Physical Review B \textbf{97}(5), 054510 (2018),
\newblock \doi{10.1103/PhysRevB.97.054510}.

\bibitem{schwarz2020}
L.~Schwarz, B.~Fauseweh, N.~Tsuji, N.~Cheng, N.~Bittner, H.~Krull, M.~Berciu,
  G.~S. Uhrig, A.~P. Schnyder, S.~Kaiser and D.~Manske,
\newblock \emph{Classification and characterization of nonequilibrium {{Higgs}}
  modes in unconventional superconductors},
\newblock Nature Communications \textbf{11}(1), 287 (2020),
\newblock \doi{10.1038/s41467-019-13763-5}.

\bibitem{dzero2024}
M.~Dzero,
\newblock \emph{Collisionless dynamics of the pairing amplitude in disordered
  superconductors},
\newblock Physical Review B \textbf{109}(10), L100503 (2024),
\newblock \doi{10.1103/PhysRevB.109.L100503}.

\bibitem{althuser2024}
J.~Alth{\"u}ser and G.~S. Uhrig,
\newblock \emph{Collective excitations in competing phases in two and three
  dimensions},
\newblock Physical Review B \textbf{109}(20), 205153 (2024),
\newblock \doi{10.1103/PhysRevB.109.205153}.

\bibitem{anderson1958}
P.~W. Anderson,
\newblock \emph{Random-{{Phase Approximation}} in the {{Theory}} of
  {{Superconductivity}}},
\newblock Physical Review \textbf{112}(6), 1900 (1958),
\newblock \doi{10.1103/PhysRev.112.1900}.

\bibitem{brieskorn1974}
G.~Brieskorn, M.~Dinter and H.~Schmidt,
\newblock \emph{Dynamics of order parameter fluctuations in gapless
  superconductors below {{{\emph{T}}}}c},
\newblock Solid State Communications \textbf{15}(4), 757 (1974),
\newblock \doi{10.1016/0038-1098(74)90255-5}.

\bibitem{simanek1975}
E.~{\v S}im{\'a}nek,
\newblock \emph{Collective modes in superconductors near {$T_c$}},
\newblock Physics Letters A \textbf{51}(4), 215 (1975),
\newblock \doi{10.1016/0375-9601(75)90535-6}.

\bibitem{schon1976}
G.~Sch{\"o}n,
\newblock \emph{Propagating Collective Modes in Superconductors},
\newblock PhD thesis, Universit{\"a}t Dortmund, Dortmund (1976).

\bibitem{maiti2015}
S.~Maiti and P.~J. Hirschfeld,
\newblock \emph{Collective modes in superconductors with competing {$s$}- and
  {$d$}-wave interactions},
\newblock Physical Review B \textbf{92}(9), 094506 (2015),
\newblock \doi{10.1103/PhysRevB.92.094506}.

\bibitem{sun2020}
Z.~Sun, M.~M. Fogler, D.~N. Basov and A.~J. Millis,
\newblock \emph{Collective modes and terahertz near-field response of
  superconductors},
\newblock Physical Review Research \textbf{2}(2), 023413 (2020),
\newblock \doi{10.1103/PhysRevResearch.2.023413}.

\bibitem{fan2022}
B.~Fan, A.~Samanta and A.~M. {Garc{\'i}a-Garc{\'i}a},
\newblock \emph{Characterization of collective excitations in weakly coupled
  disordered superconductors},
\newblock Physical Review B \textbf{105}(9), 094515 (2022),
\newblock \doi{10.1103/PhysRevB.105.094515}.

\bibitem{frohlich1952}
H.~Fr{\"o}hlich,
\newblock \emph{Interaction of electrons with lattice vibrations},
\newblock Proceedings of the Royal Society of London. Series A. Mathematical
  and Physical Sciences \textbf{215}(1122), 291 (1952),
\newblock \doi{10.1098/rspa.1952.0212}.

\bibitem{lenz1996}
P.~Lenz and F.~Wegner,
\newblock \emph{Flow equations for electron-phonon interactions},
\newblock Nuclear Physics B \textbf{482}(3), 693 (1996),
\newblock \doi{10.1016/S0550-3213(96)00521-4}.

\bibitem{mielke1997}
A.~Mielke,
\newblock \emph{Calculating critical temperatures of superconductivity from a
  renormalized {{Hamiltonian}}},
\newblock Europhysics Letters \textbf{40}(2), 195 (1997),
\newblock \doi{10.1209/epl/i1997-00445-5}.

\bibitem{mielke1997a}
A.~Mielke,
\newblock \emph{Similarity renormalization of the electron-phonon coupling},
\newblock Annalen der Physik \textbf{509}(3), 215 (1997),
\newblock \doi{10.1002/andp.19975090305}.

\bibitem{hubsch2003}
A.~H{\"u}bsch and K.~W. Becker,
\newblock \emph{Renormalization of the electron-phonon interaction: A
  reformulation of the {{BCS-gap}} equation},
\newblock The European Physical Journal B - Condensed Matter and Complex
  Systems \textbf{33}(4), 391 (2003),
\newblock \doi{10.1140/epjb/e2003-00180-9}.

\bibitem{kehrein2006}
S.~Kehrein,
\newblock \emph{The {{Flow Equation Approach}} to {{Many-Particle Systems}}},
  vol. 217 of \emph{Springer {{Tracts}} in {{Modern Physics}}},
\newblock Springer, Berlin, Heidelberg,
\newblock ISBN 978-3-540-34067-6 978-3-540-34068-3,
\newblock \doi{10.1007/3-540-34068-8} (2006).

\bibitem{krull2012}
H.~Krull, N.~A. Drescher and G.~S. Uhrig,
\newblock \emph{Enhanced perturbative continuous unitary transformations},
\newblock Physical Review B \textbf{86}(12), 125113 (2012),
\newblock \doi{10.1103/PhysRevB.86.125113}.

\bibitem{schmiedinghoff2022}
G.~Schmiedinghoff and G.~S. Uhrig,
\newblock \emph{Efficient flow equations for dissipative systems},
\newblock SciPost Physics \textbf{13}(6), 122 (2022),
\newblock \doi{10.21468/SciPostPhys.13.6.122}.

\bibitem{walther2023}
M.~R. Walther, D.-B. Hering, G.~S. Uhrig and K.~P. Schmidt,
\newblock \emph{Continuous similarity transformation for critical phenomena:
  {{Easy-axis}} antiferromagnetic {{XXZ}} model},
\newblock Physical Review Research \textbf{5}(1), 013132 (2023),
\newblock \doi{10.1103/PhysRevResearch.5.013132}.

\bibitem{hering2024}
D.-B. Hering, M.~R. Walther, K.~P. Schmidt and G.~S. Uhrig,
\newblock \emph{Quantum melting of long-range ordered quantum antiferromagnets
  investigated by momentum-space continuous similarity transformations},
\newblock Physical Review B \textbf{110}(8), 085115 (2024),
\newblock \doi{10.1103/PhysRevB.110.085115}.

\bibitem{eliashberg1960}
{\relax GM}.~Eliashberg,
\newblock \emph{Interactions between electrons and lattice vibrations in a
  superconductor},
\newblock Sov. Phys. JETP \textbf{11}(3), 696 (1960).

\bibitem{combescot1990}
R.~Combescot,
\newblock \emph{Critical temperature of superconductors: {{Exact}} solution
  from {{Eliashberg}} equations on the weak-coupling side},
\newblock Physical Review B \textbf{42}(13), 7810 (1990),
\newblock \doi{10.1103/PhysRevB.42.7810}.

\bibitem{joas2002}
C.~Joas, I.~Eremin, D.~Manske and K.~H. Bennemann,
\newblock \emph{Theory for phonon-induced superconductivity in
  {${\mathrm{MgB}}_{2}$}},
\newblock Physical Review B \textbf{65}(13), 132518 (2002),
\newblock \doi{10.1103/PhysRevB.65.132518}.

\bibitem{chang2007}
J.~Chang, I.~Eremin, P.~Thalmeier and P.~Fulde,
\newblock \emph{Eliashberg theory of superconductivity and inelastic rare-earth
  impurity scattering in the filled skutterudite
  {${\mathrm{La}}_{1\ensuremath{-}x}{\mathrm{Pr}}_{x}{\mathrm{Os}}_{4}{\mathrm{Sb}}_{12}$}},
\newblock Physical Review B \textbf{76}(22), 220510 (2007),
\newblock \doi{10.1103/PhysRevB.76.220510}.

\bibitem{morel1962}
P.~Morel and P.~W. Anderson,
\newblock \emph{Calculation of the {{Superconducting State Parameters}} with
  {{Retarded Electron-Phonon Interaction}}},
\newblock Physical Review \textbf{125}(4), 1263 (1962),
\newblock \doi{10.1103/PhysRev.125.1263}.

\bibitem{carbotte1990}
J.~P. Carbotte,
\newblock \emph{Properties of boson-exchange superconductors},
\newblock Reviews of Modern Physics \textbf{62}(4), 1027 (1990),
\newblock \doi{10.1103/RevModPhys.62.1027}.

\bibitem{sigrist2005}
M.~Sigrist,
\newblock \emph{Introduction to {{Unconventional Superconductivity}}},
\newblock AIP Conference Proceedings \textbf{789}(1), 165 (2005),
\newblock \doi{10.1063/1.2080350}.

\bibitem{kalthoff2017}
M.~Kalthoff, F.~Keim, H.~Krull and G.~S. Uhrig,
\newblock \emph{Comparison of the iterated equation of motion approach and the
  density matrix formalism for the quantum {{Rabi}} model},
\newblock The European Physical Journal B \textbf{90}(5), 97 (2017),
\newblock \doi{10.1140/epjb/e2017-80063-2}.

\bibitem{uhrig2009}
G.~S. Uhrig,
\newblock \emph{Interaction quenches of {{Fermi}} gases},
\newblock Physical Review A \textbf{80}(6), 061602 (2009),
\newblock \doi{10.1103/PhysRevA.80.061602}.

\bibitem{hamerla2013}
S.~A. Hamerla and G.~S. Uhrig,
\newblock \emph{Dynamical transition in interaction quenches of the
  one-dimensional {{Hubbard}} model},
\newblock Physical Review B \textbf{87}(6), 064304 (2013),
\newblock \doi{10.1103/PhysRevB.87.064304}.

\bibitem{hamerla2014}
S.~A. Hamerla and G.~S. Uhrig,
\newblock \emph{Interaction quenches in the two-dimensional fermionic
  {{Hubbard}} model},
\newblock Physical Review B \textbf{89}(10), 104301 (2014),
\newblock \doi{10.1103/PhysRevB.89.104301}.

\bibitem{rickayzen1980}
G.~Rickayzen,
\newblock \emph{Green's Functions and Condensed Matter},
\newblock No.~5 in Techniques of Physics. Academic Press, London,
\newblock ISBN 978-0-12-587950-7 (1980).

\bibitem{czycholl2008}
G.~Czycholl,
\newblock \emph{{Theoretische Festk{\"o}rperphysik}},
\newblock {Springer-Lehrbuch}. Springer, Berlin, Heidelberg,
\newblock ISBN 978-3-540-74789-5 978-3-540-74790-1,
\newblock \doi{10.1007/978-3-540-74790-1} (2008).

\bibitem{simonato2023}
M.~Simonato, M.~I. Katsnelson and M.~R{\"o}sner,
\newblock \emph{Revised {{Tolmachev-Morel-Anderson}} pseudopotential for
  layered conventional superconductors with nonlocal {{Coulomb}} interaction},
\newblock Physical Review B \textbf{108}(6), 064513 (2023),
\newblock \doi{10.1103/PhysRevB.108.064513}.

\bibitem{tolmachev1961}
V.~V. Tolmachev,
\newblock \emph{Logarithmic criterion for superconductivity},
\newblock Dokl. Akad. Nauk SSSR \textbf{140}, 563 (1961).

\bibitem{kittel2004}
C.~Kittel,
\newblock \emph{Introduction to {{Solid State Physics}}},
\newblock John Wiley \& Sons, New York, 8 edn.,
\newblock ISBN 978-0-471-68057-4 (2004).

\bibitem{kittel1963}
C.~Kittel,
\newblock \emph{Quantum {{Theory}} of {{Solids}}},
\newblock John Wiley \& Sons, New York, 1 edn.,
\newblock ISBN 0-471-49025-3 (1963).

\bibitem{vashishta1973}
P.~Vashishta and J.~P. Carbotte,
\newblock \emph{Superconductivity in
  {${\mathrm{Pb}}_{0.9}$}{${\mathrm{Bi}}_{0.1}$}},
\newblock Physical Review B \textbf{7}(5), 1874 (1973),
\newblock \doi{10.1103/PhysRevB.7.1874}.

\bibitem{kostrzewa2018}
M.~Kostrzewa, R.~Szcz{\k e}{\'s}niak, J.~K. Kalaga and I.~A. Wrona,
\newblock \emph{Anomalously high value of {{Coulomb}} pseudopotential for the
  {{H5S2}} superconductor},
\newblock Scientific Reports \textbf{8}(1), 11957 (2018),
\newblock \doi{10.1038/s41598-018-30391-z}.

\bibitem{pettifor1984}
D.~Pettifor and D.~Weaire,
\newblock \emph{The {{Recursion Method}} and {{Its Applications}}},
\newblock Springer, Berlin, Heidelberg,
\newblock ISBN 978-3-642-82444-9 (1984).

\bibitem{viswanath1994}
V.~Viswanath and G.~M{\"u}ller,
\newblock \emph{The {{Recursion Method}}},
\newblock Springer, Berlin, Heidelberg,
\newblock ISBN 978-3-662-14512-8 (1994).

\bibitem{goldstone1961}
J.~Goldstone,
\newblock \emph{Field theories with << {{Superconductor}} >> solutions},
\newblock Il Nuovo Cimento (1955-1965) \textbf{19}(1), 154 (1961),
\newblock \doi{10.1007/BF02812722}.

\bibitem{anderson1963}
P.~W. Anderson,
\newblock \emph{Plasmons, {{Gauge Invariance}}, and {{Mass}}},
\newblock Physical Review \textbf{130}(1), 439 (1963),
\newblock \doi{10.1103/PhysRev.130.439}.

\bibitem{barankov2007}
R.~A. Barankov and L.~S. Levitov,
\newblock \emph{Excitation of the dissipationless {{Higgs}} mode in a fermionic
  condensate} (arXiv:0704.1292) (2007),
\newblock \doi{10.48550/arXiv.0704.1292},
\newblock \eprint{0704.1292}.

\bibitem{coleman2015}
P.~Coleman,
\newblock \emph{Introduction to {{Many-Body Physics}}},
\newblock Cambridge University Press, Cambridge,
\newblock ISBN 978-0-521-86488-6,
\newblock \doi{10.1017/CBO9781139020916} (2015).

\bibitem{hirashima1987}
D.~S. Hirashima and H.~Namaizawa,
\newblock \emph{Collective {{Excitations}} in p- and d-{{Wave
  Superconductors}}},
\newblock Japanese Journal of Applied Physics \textbf{26}(S3-1), 167 (1987),
\newblock \doi{10.7567/JJAPS.26S3.167}.

\bibitem{bardasis1961}
A.~Bardasis and J.~R. Schrieffer,
\newblock \emph{Excitons and {{Plasmons}} in {{Superconductors}}},
\newblock Physical Review \textbf{121}(4), 1050 (1961),
\newblock \doi{10.1103/PhysRev.121.1050}.

\bibitem{bohm2014}
T.~B{\"o}hm, A.~F. Kemper, B.~Moritz, F.~Kretzschmar, B.~Muschler, H.-M. Eiter,
  R.~Hackl, T.~P. Devereaux, D.~J. Scalapino and H.-H. Wen,
\newblock \emph{Balancing {{Act}}: {{Evidence}} for a {{Strong Subdominant}}
  {$d$}-{{Wave Pairing Channel}} in
  {${\mathrm{Ba}}_{0.6}{\mathrm{K}}_{0.4}{\mathrm{Fe}}_{2}{\mathrm{As}}_{2}$}},
\newblock Physical Review X \textbf{4}(4), 041046 (2014),
\newblock \doi{10.1103/PhysRevX.4.041046}.

\bibitem{sun2020a}
Z.~Sun and A.~J. Millis,
\newblock \emph{Bardasis-{{Schrieffer}} polaritons in excitonic insulators},
\newblock Physical Review B \textbf{102}(4), 041110 (2020),
\newblock \doi{10.1103/PhysRevB.102.041110}.

\bibitem{muller2021}
M.~A. M{\"u}ller and I.~M. Eremin,
\newblock \emph{Signatures of {{Bardasis-Schrieffer}} mode excitation in
  third-harmonic generated currents},
\newblock Physical Review B \textbf{104}(14), 144508 (2021),
\newblock \doi{10.1103/PhysRevB.104.144508}.

\bibitem{hackner2023}
N.~A. Hackner and P.~M.~R. Brydon,
\newblock \emph{Bardasis-{{Schrieffer-like}} phase mode in a superconducting
  bilayer},
\newblock Physical Review B \textbf{108}(22), L220505 (2023),
\newblock \doi{10.1103/PhysRevB.108.L220505}.

\end{thebibliography}

%%%%%%%%%% END TODO: BIBLIOGRAPHY

\end{document}